\documentclass[useAMS,usenatbib]{mn2e}

\usepackage{graphicx}
\def\mnras{MNRAS}
\def\aap{A\&A}

\def\apj{APJ}

\def\aj{AJ}

\newcommand{\msun}{\mbox{M$_{\odot}$}}

\newcommand{\kms}{\mbox{$\rm{\,km\,s^{-1}}$}}

\begin{document}

\title[SN~2012ec]{SN 2012ec: mass of the progenitor from PESSTO follow-up of the photospheric phase.}
\author[C. Barbarino et al. ]{C. Barbarino$^{1,2}$, M.Dall'Ora$^{2}$, M.T. Botticella$^{2}$, M. Della Valle$^{2,3}$, L. Zampieri $^{4}$,
\newauthor J.R. Maund$^{5}$, M.L. Pumo$^{4}$ , A. Jerkstrand$^{6}$, S. Benetti$^{4}$, N. Elias-Rosa$^{4}$, M. Fraser$^{17}$,
\newauthor A. Gal-Yam$^{7}$, M. Hamuy$^{15,16}$, C. Inserra$^{6}$, C. Knapic$^{8}$, A.P. LaCluyze$^{14}$, M. Molinaro$^{8}$,
\newauthor P. Ochner$^{4}$, A. Pastorello $^{4}$, G. Pignata$^{9,15}$, D.E. Reichart$^{14}$, C. Ries$^{12}$, A. Riffeser$^{12}$,
\newauthor  B. Schmidt$^{10}$, M. Schmidt$^{12}$, R. Smareglia$^{8}$, S.J. Smartt$^{6}$, K. Smith$^{6}$, J. Sollerman$^{18}$,
\newauthor M. Sullivan$^{11}$, L. Tomasella$^{4}$,M. Turatto$^{4}$, S. Valenti$^{13}$, O. Yaron$^{7}$ and D. Young$^{6}$. \\
$^{1}$Dip. di Fisica and ICRA, Sapienza Universit\'{a} di Roma, Piazzale Aldo Moro 5, I-00185 Rome, Italy\\
$^{2}$INAF- Osservatorio Astronomico di Capodimonte, Salita Moiariello 16, 80131 Napoli, Italy \\
$^{3}$ICRANet-Pescara, Piazza della Repubblica 10, I-65122 Pescara, Italy \\
$^{4}$INAF- Osservatorio Astronomico di Padova, Vicolo dell'Osservatorio 5, 35122 Padova, Italy \\
$^{5}$Department of Physics and Astronomy, University of Sheffield, Sheffield, S3 7RH, United Kingdom\\
$^{6}$Astrophysics Research Centre, School of Mathematics and Physics,  Queen's University Belfast, Belfast BT7 1NN, United Kingdom\\
$^{7}$Department of Particle Physics and Astrophysics, The Wiezmann Institute of Science, Rehovot, 76100 Israel \\
$^{8}$INAF- Osservatorio Astronomico di Trieste, Via Tiepolo Giambattista 11, 34131, Trieste, Italy \\
$^{9}$Departemento de Ciencias Fisicas, Universidad Andres Bello, Avda Republica 252, Santiago, Chile  \\
$^{10}$Research School of Astronomy and Astrophysics, Australian National University, Cotter Road, Weston Creek, ACT 2611, Australia \\
$^{11}$School of Physics and Astronomy, University of Southampton, Southampton, SO17 1BJ, UK \\
$^{12}$Universit\"{a}ts-Sternwarte M\"{u}nchen, Scheinerstr. 1, D-81679 M\"{u}nchen, Germany\\
$^{13}$Las Cumbres Observatory Global Telescope Network, 6740 Cortona Dr., Suite 102, Goleta, CA 93117, USA \\
$^{14}$Department of Physics and Astronomy, University of North Carolina at Chapel Hill, Chapel Hill, NC 27599-3255, USA \\
$^{15}$Millennium Institute of Astrophysics (MAS), Santiago, Chile \\
$^{16}$Departamento de Astronomia, Universidad de Chile, Casilla 36-D, Santiago, Chile \\
$^{17}$Institute of Astronomy, University of Cambridge, Madingley Road, Cambridge, CB3 0HA, UK \\
$^{18}$The Oskar Klein Centre, Department of Astronomy, Stockholm University, Albanova, 10691 Stockholm, Sweden
}

\maketitle

\begin{abstract}

We present the results of a photometric and spectroscopic monitoring campaign of SN 2012ec, which exploded in the spiral galaxy NGC 1084, during the photospheric phase. The photometric light curve exhibits a plateau with luminosity $\mathrm{L= 0.9 \times 10^{42} \: erg \: s^{-1}}$ and duration $\sim $90 days, which is somewhat shorter than standard Type II-P supernovae.  We estimate the nickel mass $\mathrm{M(^{56}Ni)= 0.040 \pm 0.015 \: \msun}$ from the luminosity at the beginning of the radioactive tail of the light curve.
The explosion parameters of SN 2012ec were estimated from the comparison of the bolometric light curve and the observed temperature and velocity evolution of the ejecta with predictions from hydrodynamical models.
We derived an envelope mass of $12.6 \: \msun$, an initial progenitor radius of $\mathrm{1.6 \times 10^{13} \: cm}$ and an explosion energy of $\mathrm{1.2 \: foe}$.
These estimates agree with an independent study of the progenitor star identified in pre-explosion images, for which an initial mass of $\mathrm{M=14-22 \: \msun}$ was determined.
We have applied the same analysis to two other type II-P supernovae (SNe 2012aw and 2012A), and carried out a comparison with the properties of SN 2012ec derived in this paper. We find a reasonable agreement between the masses of the progenitors obtained from pre-explosion images and masses derived from hydrodynamical models. We estimate the distance to SN 2012ec with the Standardized Candle Method (SCM) and compare it with other estimates based on other primary and secondary indicators. SNe 2012A, 2012aw and 2012ec all follow the standard relations for the SCM for the use of Type II-P SNe as distance indicators.

\end{abstract}  


\section{Introduction}\label{Introduction}

Core-collapse supernovae (CC-SNe) originate from the gravitational collapse of the iron cores formed by massive stars ($M \geq 8 \: \msun$) that cannot be supported by further exothermal thermonuclear reactions (\citealt{Iben1983}; \citealt{Woosley2002}).  An important sub-class of CC-SNe is represented by Type II-plateau events (SNe II-P) characterized by the presence of hydrogen in their spectra \citep{Filippenko1997} and a luminosity ``plateau'' that lasts for $\sim 80 - 100$ days, after the blue band maximum of the light curve \citep{Barbon1979}. The plateau is powered by the recombination of hydrogen in the SN ejecta.  When the recombination ends, the lightcurve drops sharply by several magnitudes in $\sim 30$ days (e.g. \citealt{Kasen2009}; \citealt{Olivares2010}).
This transition phase is followed by a linear ``radioactive tail'', where the light curve is powered by the radioactive decay of $^{56}$Co to $^{56}$Fe.  In this phase the SN luminosity depends on the amount of $^{56}$Ni synthesized in the explosion (e.g. \citealt{Weaver1980}).\\

Both theoretical (e.g. \citealt{Grassberg1971}; \citealt{Litvinova1983}; \citealt{Utrobin2008}; \citealt{Pumo2011}; \citealt{Bersten2012}) and empirical (e.g. \citealt{Smarttetal2009}) investigations show that type II-P SNe are generally associated with red supergiants (RSGs). A minor fraction of them (less than $3-5\%$, e.g. \citealt{Smarttetal2009}; \citealt{Kleiser2011}; \citealt{Pastorello2012}) results from the explosion of a blue supergiant, similar to SN 1987A (\citealt{Gilmozzi1987}; \citealt{Kirshner1987}). Theoretical models predict that type II-P SNe are the final fate of progenitors between $8$ and $30$ $M_\odot $ (e.g. \citealt{Heger2003}; \citealt{Walmswell2012}).
Most progenitors identified in high-resolution archival images were found to be RSGs of initial masses between $\sim 8 \: \msun$ and $\sim 17 \: \msun$.  The apparent lack of high-mass progenitors has been dubbed as the ``RSG problem'' (\citealt{Smarttetal2009}, and references therein).
The existence of this discrepancy has been further confirmed by studies of the massive star population in Local Group galaxies, for which RSGs have been found to have masses up to $25 \: \msun$ (\citealt{Massey2000}; \citealt{Massey2001}).\\

The reason for this lack of detection of massive RSG progenitors is still debated.  A possible solution of the RSG problem was presented by \citet{Walmswell2012}. They speculate that an underestimation of the luminosity of the RSG SN progenitors (and therefore of their masses) might occur if we neglect the presence of an additional extinction due to dust production in the RSG winds. They estimated a new upper limit for the mass range of $21^{+2}_{-1} \msun$, which is, within the errors, marginally consistent with the range derived by \citet{Smartt2009}. \citet{Kochanek2012} pointed out that the use of standard interstellar extinction laws may overestimate the effects of the reddening.\\

A different approach to estimate the mass of Type II-P SN progenitors is based on the use of hydrodynamic modelling of the SN evolution. This allows us to determine the ejecta mass, explosion energy, pre-SN radius and Ni mass by performing a simultaneous comparison between the observed and simulated light curves, the evolution of line velocities and the continuum temperature (\citealt{Litvinova1983}; \citealt{Litvinova1985}; \citealt{Zampieri2005}; \citealt{Zampieri2007}).
The pre-explosion mass is calculated from the ejecta mass assuming the mass of a neutron star remnant ($1.4 \: \msun$) and mass loss through stellar winds. The hydrodynamic modelling of several well-observed Type II-P SNe (SNe 1997D, \citealt{Zampieri1998}; 1999em, \citealt{Elmhamdi2003}; 2003Z, \citealt{Utrobin2007} and \citealt{Spiro2014}; 2004et, \citealt{Maguire2010}; 2005cs, \citealt{Pastorello2009}; 2009kf, \citealt{Botticella2010}) determined higher masses for the progenitors than those derived from the analysis of pre-explosion images. This discrepancy either points to systematic errors in the analysis of pre-explosion images or in the assumptions in the physiscs of the hydrodinamical modelling (\citealt{Utrobin1993}; \citealt{Blinnikov2000};  \citealt{Chugai2000}; \citealt{Zampieri2003}; \citealt{Pastorello2004}; \citealt{Utrobin2007a}, \citealt{Utrobin2007}; \citealt{Utrobin2008}; \citealt{Utrobin2009}; \citealt{Pastorello2009b}).\\

Another method to estimate the mass of the progenitor is the modeling of nebular phase spectroscopic observations (\citealt{Jerkstrand2012}; \citealt{Jerkstrand2014}) ,which provide good agreement with estimates obtained by the analysis of pre-explosion images.\\

The astrophysical interest in Type II-P SNe is twofold: 1) observations show that Type II-P SNe are the most common explosions in the nearby Universe (e.g. \citealt{Cappellaro1999}; \citealt{Li2011}); and 2) starting from the pioneering suggestion by \citet{Kirshner1974}, Type II-P SNe have been proposed as robust distance indicators.
Two different approaches are used to derive distance measurements of SNe II-P.
The  theoretical approach is  based on spectral modelling like the expanding photosphere method (e.g. \citealt{Eastman1996}) or the spectral expanding atmosphere method (e.g., \citealt{Baron2004}). 
Empirical approaches exploit  the observed correlation  between
the luminosity of a Type II-P SN and its expansion velocity (e.g., the standardized candle method, \citealt{Hamuy2002}) or the steepness of the light curve after the plateau phase \citep{Elmhamdi2003b}.
The \citet{Hamuy2002} method, refined for example by \citet{Nugent2006}, \citet{Poznanski2009}, and \citet{Olivares2010}, has an intrinsic accuracy of $\sim 10-12\%$ \citep{Hamuy2002}; slightly larger than the accuracy obtained for Type Ia SNe (e.g. \citealt{Tammann2013}). Type II-P SNe can, importantly, be observed out to cosmological distances (e.g. \citealt{Nugent2006}); with the advantage of being that they arise from a homogenous progenitor population.  The \citet{Hamuy2002} method can, therefore, be used as an independent health check of the SN Ia-based distance scale.

The main goal of this paper is to present the results of our photometric and spectroscopic monitoring campaign of SN 2012ec, which exploded in NGC 1084. The early data were collected via the Large Program ``Supernova Variety and Nuclesosynthesis Yelds''(PI S. Benetti). A substantial fraction of the data has been collected via the ESO Public Survey PESSTO \footnote{www.pessto.org} (``Public ESO Spectroscopic Survey of Transient Objects'', PI S.J. Smartt). The observations of SN 2012ec were analysed in conjunction with the hydrodynamical codes described in \citet{Pumo2010} and \citet{Pumo2011}, and information on the progenitor obtained from high-resolution pre-explosion images. The same analysis has already performed for two other type II-P SNe: SN 2012A (\citealt{Tomasella2013}; \citealt{Roy2014}) and SN 2012aw (\citealt{Fraser2012}; \citealt{Bayless2013}; \citealt{Bose2013}; \citealt{DallOra2014}). This allows us to carry out an homogeneous comparative study of these three SNe, and to identify possible systematic discrepancies in the estimates of the masses of the progenitors derived from different techniques.

The paper is organized as follows: in Section 2 we present the discovery and the detection of the progenitor of SN 2012ec; in Section 3 we discuss the properties of the host galaxy, the distance and the extinction; in Section 4 we present the optical and near-infrared (NIR) photometric evolution of SN 2012ec, and compare its colour evolution and bolometric light curve with those of other Type II-P SNe. In Section 5 we present the optical and NIR spectroscopic observations. In Section 6 we discuss the results of the modeling of the data and in Section 7 we present a detailed comparison of SN 2012ec with the Type II-P SNe 2012A and 2012aw.  In Section 8 we consider these three SNe in the context of the SCM and in Section 9 we discuss our results.

\section{Discovery and progenitor detection}\label{discovery}

\begin{figure}
  \begin{center}
  \includegraphics[scale=0.4]{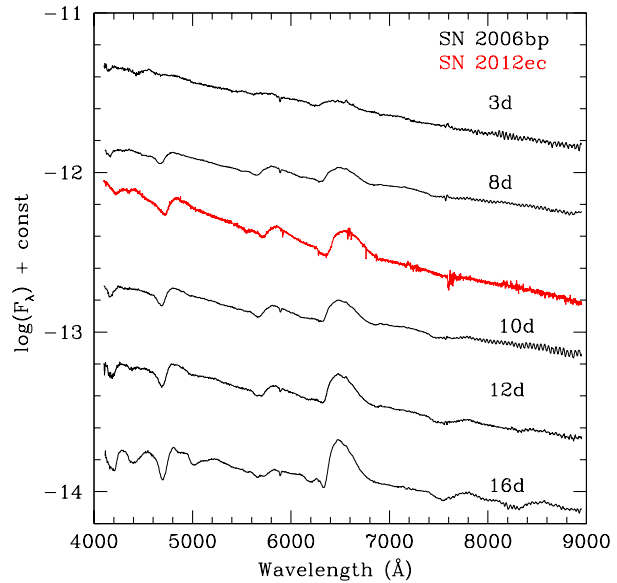}
  \end{center}
  \caption{Comparison between a very early spectrum of SN 2012ec and $5$ spectra of SN 2006bp, from day $3$ to $16$.}
  \label{comp}
\end{figure}

SN 2012ec was discovered by \citet{Monard2012} in the almost face-on ($i=57^{\circ}$, \citealt{Moiseev2000}) spiral galaxy NGC 1084 on 2012 August 11.039 UT (MJD=56150.04). \citet{Childress2012} classified SN 2012ec as a very young type II-P SN, probably a few days post-explosion. In Fig. \ref{comp} we show this early spectrum of SN 2012ec (collected on 2012, August 13 with WiFeS, MJD $= 56152.2$), compared with SN 2006bp \citep{Quimby2007} at five different epochs.  The spectrum of SN 2012ec is very similar to those of SN 2006bp \citep{Quimby2007} obtained at $8$ and $10$ days after the explosion, implying that the SN was observed at $\sim +9$ days post-explosion and giving an explosion epoch of $\sim 7$ days before the discovery. We explicitly note that our estimate is slightly different from the one given by \citet{Maund2013}, who estimated the explosion date at $<6$ days before the discovery by comparison with spectra of SN~1999em.
The explosion epoch of SN 2006bp is much more tightly constrained than that of SN 1999em, because it is based on the detection of shock breakout (\citealt{Nakano2006}; \citealt{Quimby2007}). The estimates obtained by using either SN 2006bp or SN 1999em, as reference, are in agreement within the errors. We adopt, therefore, a conservative constraint on the explosion date of $7 \pm 2$ days prior to discovery and define the zero phase as our estimated explosion epoch of MJD $= 56143.0$.

\citet{Maund2013} identified a progenitor candidate in pre-explosion Hubble Space Telescope (HST) images.  Photometry of the progenitor candidate was compared with synthetic photometry of MARCS spectral energy distributions (SED) \citep{Gustafsson2008}, which suggested that the progenitor of SN 2012ec was a RSG with an initial mass in the range $14-22 \: \msun$.

\section{Host galaxy, distance and extinction}\label{hostgalsec}

The SN is located $0.7"$E and $15.9"$N of the nucleus of the host galaxy NGC 1084 (see Fig \ref{FC}).  Details of NGC~1084 are presented in Table \ref{galprop}.  NGC 1084 previously hosted $4$ known SNe: the Type II-P SN 2009H \citep{Li2009}, the Type II SNe 1998dl \citep{King1998} and 1996an \citep{Nakano1996}, and the Type Ia SN 1963P \citep{Kowal1968}.

The distances available in the literature for NGC 1084 are principally based on the Tully-Fisher relation, and we adopt the value $\mu = 31.19 \pm 0.13$ mag, available in the Extragalactic Distance Database \footnote{Extragalactic Distance Database, \hspace*{0.16cm} http:/$\!$/edd.ifa.hawaii.edu/} \citep{Tully2009}.

The Galactic reddening towards SN~2012ec was estimated from the \citet{Schlafly2011} dust maps to be $E(B-V)= 0.024$ mag\footnote{We checked the consistency with the \citet{Schlegel1998} calibration, and they agree within a few thousandths of magnitude} .  The internal reddening in NGC~1084 was derived using the measured equivalent widths (EW) of NaI D ($5889$, $5895$  \AA), observed in a low-resolution spectrum at $+19$ days. The measured value was $\mathrm{EW(NaID) = 0.8 \pm 0.3}$ \AA\, from which we obtained $E(B-V)= 0.12 ^{+0.15}_{-0.12}$ mag using the \citet{Poznanski2012} calibration and $E(B-V)=0.11$ mag using the \citet{Turatto2003} calibration. These  two values are in good agreement and we adopt $E(B-V)=0.12 ^{+0.15}_{-0.12}$ mag for the host galaxy reddening. 

Assuming a \citet{Cardelli1989} reddening law ($R_{V}=3.1$), we estimate the total Galactic and host $V-$band extinction towards SN~2012ec to be $A_{V}= 0.45$ mag.

\begin{table}
\caption{Properties of NGC 1084.\label{galprop}}
\begin{footnotesize}
\begin{tabular}{ll}
\hline
$\alpha$ (2000) & $2^{h}43^{m}32.091$  \\					 
$\delta$ (2000) & $-07\degr 47\arcmin 16.76\arcsec$ \\
morphological type &  SA(s)d  \\
\texttt{z} & $0.004693 \pm 0.000013$  \\
$\mu$ & $31.19 \pm 0.13$ mag \\
v$_{Hel}$  & $1407 \pm 4 \kms$  \\
$E(B-V)_{Galactic}$ & $0.024$ mag  \\
$E(B-V)_{host}$ & $0.12$ mag \\
\hline
\end{tabular}
\\[1.5ex]
\end{footnotesize}
\end{table}

\section{Photometric evolution}\label{Photsec}

\subsection{Data sample and reduction}

A photometric and spectroscopic monitoring campaign for SN~2012ec, at optical and NIR wavelengths, was conducted over a period $153$ days, covering $77$ epochs from $11$ to $164$ days post-explosion, using multiple observing facilities.  Additional data collected in the nebular phase will be published in a companion paper (\citealt{Jerkstrand14b}, subm.).

$BVRI$ Johnson-Cousins data were collected with: the $2.0$m Liverpool Telescope (LT, Canary Islands, Spain) equipped with the IO:O camera ($BV$, $21$ epochs); the $3.58$m ESO New Technology Telescope (NTT, La Silla, Chile) equipped with the EFOSC2 (ESO Faint Object and Spectrograph Camera) camera ($BVRI$, $9$ epochs); the $1.82$m Copernico telescope (Asiago, Italy) equipped with the AFOSC Asiago Faint Object Spectrograph and Camera ($BVRI$; $3$ epochs); the $0.6$m ESO TRAnsiting Planets and PlanetesImals Small Telescope (TRAPPIST, La Silla, Chile), equipped with TRAPPISTCAM ($BVR$, $4$ epochs); and the the array of $0.41$m Panchromatic Robotic Optical Monitoring and Polarimetry Telescopes (PROMPT, Cerro Tololo, Chile), equipped with Apogee U47p cameras, which employ the E2V CCDs ($BVRI$, $21$ epochs).

$ugriz$ images were collected with: the LT equipped with the IO:O camera ($uriz$ $21$ epochs); the ESO NTT Telescope equipped with EFOSC2 ($ugriz$, $3$ epochs); the PROMPT telescopes ($griz$, $19$ epochs); and the $0.4$m telescope at the Wendelstein Observatory (Mount Wendelstein, Germany), equipped with a ST-10 CCD camera ($gri$, $7$ epochs). 

$JHK_{s}$ observations were acquired with the ESO NTT telescope, equipped with the SOFI (Son Of ISAAC) camera ($8$ epochs).

A summary of the characteristics of the instruments and telescopes used for photometric follow up are presented in Table \ref{phtel}.

\begin{table*}
\caption{Summary of the characteristics of the instruments used during for photometric monitoring.\label{phtel}}
\begin{footnotesize}
\begin{tabular}{lclllcll}
\hline
Telescope  &Camera & Pixel scale & Field of view & Filters$^{a}$ & \# of epochs \\
 &  & [arcsec/pix] & [arcmin] &   \\
\hline
NTT (3.58m)      & EFOSC2          & 0.24    & 4 $\times$ 4     & $B, V, R$; $u, g,r, i$          & 12 \\
NTT (3.58m)      & SOFI            & 0.28    & 5 $\times$ 5     & $J, H, K_{s}$              &  8 \\
LT  (2.0m)       & IO:O            & 0.15    & 10 $\times$ 10   & $B, V$: $u, r, i, z$   & 21 \\
PROMPT (0.41m)   & APU9            & 0.59    & 11 $\times$ 11   & $B, V, R, I$; $g, r, i, z$ & 21 \\
CAO    (1.82m)    & AFOSC           & 0.46    & 8 $\times$ 8     & $B, V, R$; $i$          &  3 \\
SAO    (0.97m)   & SBIG            & 0.86    & 57 $\times$ 38   & $R$                        &  1 \\
WOT    (0.4m)    & SBIG ST-10 XME  & 0.44    & 16 $\times$ 10   & $g, r, i$          &  7 \\
TRAPPIST (0.60m) & TRAPPISTCAM     & 0.65    & 27 $\times$ 27   & $B$, $V$, $R$              &  4 \\
\hline
\end{tabular}
\\[1.5ex]
NTT = New Technology Telescope with the optical camera ESO Faint Object Spectrograph and Camera EFOSC2 and with the Near-Infrared Camera Son of ISAAC (SOFI); LT = the Liverpool Telescope (LT) with the optical CCD CAMERA IO:O; PROMPT = Panchromatic Robotic Optical Monitoring and Polarimetry Telescopes; CAO = the Copernico telescope at Asiago Observatory with the Asiago Faint Object Spectrograph and Camera (AFOSC); SAO = the Schmidt telescope at the Asiago Observatory; WOT = the $40~cm$ telescope at the Wendelstein Observatory; TRAPPIST = TRAnsit Planets and PlanetesImals Small Telescope. \\
$^{a}$ The NTT and CAO i filter is Gunn.

\end{footnotesize}
\end{table*}

Data were pre-reduced using the respective instrument pipelines, where available, or following the standard procedures (bias, overscan and flat-field corrections, trimming) in the \texttt{IRAF} \footnote{IRAF is distributed by the National Optical Astronomical Observatory, which is operated by the Association of Universities for Research in Astronomy, Inc., under cooperative agreement with the National Science Foundation.} environment. In particular, NIR images were pre-reduced by means of an \texttt{IRAF}-based custom pipeline based on the \texttt{XDIMSUM IRAF} package \citep{Coppola2011}, which conducts the background subtraction using a two-step technique based on a preliminary guess of the sky background and with a careful masking of unwanted sources in the sky images.

Johnson-Cousins $BVRI$ calibrated magnitudes of $18$ reference stars were obtained by averaging their photometry obtained on $12$ photometric nights, in conjunction with observations of  \citet{Landolt1992} standard star fields. $ugriz$ calibrated photometry for $17$ reference stars were obtained on $11$ photometric nights with the LT and the NTT telescopes, in conjunction with observations of \citet{Smith2002} $u'g'r'i'z'$ standard star fields. Finally, calibrated NIR 2MASS $JHK$ photometry was obtained for $5$ reference stars, for which 2MASS \citep{Skrutskie2006} photometry was available. We did not correct NIR magnitudes for colour terms, since they are generally very small in the NIR bands (e.g. \citealt{Carpenter2001}). Our adopted reference stars showed no clear signs of variability.

\begin{figure}
  \centering
  \includegraphics[scale=0.25]{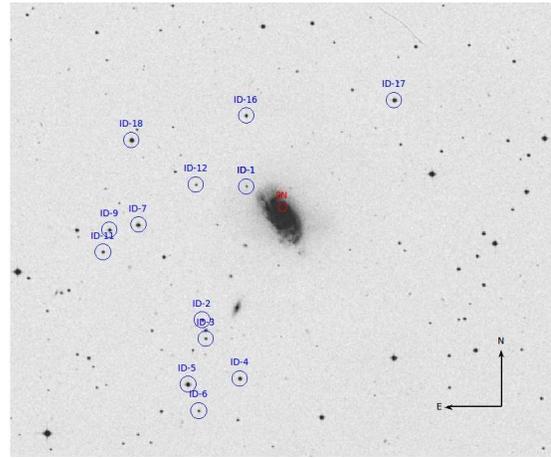}
  \caption{An image of SN 2012ec and the host galaxy NGC 1084, acquired with the Liverpool Telescope and the IO:O camera. The field of view is $14.5 \times 14.5$ arcmin$^2$.  Reference stars are circled and labeled (see Tables \ref{local} and \ref{local1}).}
  \label{FC}
\end{figure}

The host galaxy and the SN position are shown in Fig. \ref{FC}, along with the local sequence stars adopted for the photometric calibration. The calibrated photometry for the local sequence stars is reported in Tables \ref{local} and \ref{local1}. In the following, the Johnson-Cousins $BVRI$ and NIR photometry are reported in Vega magnitudes, while the $ugriz$ photometry is reported in the AB magnitude system.

\begin{table*}
\setlength{\tabcolsep}{3pt}.
\caption{Positions and photometry of the local sequence reference stars in the $BVRI$ and in the $u'g'r'i'z'$ systems.\label{local}}
\begin{scriptsize}
\begin{tabular}{llllllllllll}
\hline
 \# id & $ \alpha_{J2000.0} $ & $ \delta_{J2000.0} $ & $B$ & $V$  & $R$  & $I$ &$u'$ &$g'$ &$r'$ &$i'$ &$z'$ \\
 & (deg) &(deg) & mag & mag & mag & mag & mag & mag & mag & mag & mag \\
\hline 
1  & 41.5216674 & -7.5597940 &  17.98 (0.02) & 16.88 (0.02) & 16.19 (0.02) & 15.53 (0.03) & 19.85 (0.02) & 17.48 (0.04) & 16.41 (0.01) & 16.01 (0.02) & 15.86 (0.01) \\
2  & 41.5496917 & -7.6416869 &  16.84 (0.02) & 15.97 (0.02) & 15.46 (0.02) & 14.93 (0.03) & 18.27 (0.02) & 16.43 (0.02) & 15.67 (0.02) & 15.38 (0.01) & 15.32 (0.02) \\
3  & 41.5474764 & -7.6530580 &  17.14 (0.02) & 16.28 (0.02) & 15.81 (0.02) & 15.32 (0.02) & 18.45 (0.06) & 16.71 (0.03) & 16.04 (0.02) & 15.78 (0.01) & 15.72 (0.01) \\
4  & 41.5265649 & -7.6778087 &  15.58 (0.02) & 14.95 (0.02) & 14.66 (0.02) &              & 16.48 (0.03) & 15.25 (0.02) & 14.80 (0.02) & 14.64 (0.02) & 14.65 (0.01) \\
5  & 41.5589242 & -7.6811761 &  14.27 (0.02) & 13.52 (0.02) & 13.23 (0.02) &              & 15.18 (0.02) & 13.91 (0.02) & 13.31 (0.02) & 13.05 (0.02) & 12.94 (0.01) \\
6  & 41.5522025 & -7.6973300 &  17.01 (0.02) & 16.00 (0.02) & 15.43 (0.02) &              & 18.82 (0.05) & 16.55 (0.03) & 15.62 (0.02) & 15.30 (0.02) & 15.18 (0.01) \\
7  & 41.5886692 & -7.5829251 & 14.62 (0.01) & 14.07 (0.01) & 13.84 (0.02) &              & 15.36 (0.03) & 14.49 (0.02) & 13.98 (0.01) & 13.86 (0.02) & 13.87 (0.01) \\
9  & 41.6065545 & -7.5858940 & 15.16 (0.01) & 14.42 (0.01) & 14.08 (0.02) &              & 16.11 (0.02) & 15.03 (0.03) & 14.25 (0.02) & 14.04 (0.02) & 14.01 (0.01) \\
11 & 41.6108998 & -7.5996059 & 16.99 (0.01) & 16.36 (0.01) & 16.04 (0.02) &              & 17.64 (0.02) & 16.66 (0.03) & 16.21 (0.01) & 15.99 (0.02) & 15.97 (0.02) \\
12 & 41.5528922 & -7.5585795 & 18.32 (0.02) & 16.83 (0.02) & 15.93 (0.02) & 14.73 (0.02) & 20.12 (0.05) & 17.65 (0.01) & 16.19 (0.01) & 15.24 (0.02) & 14.90 (0.02) \\
16 & 41.5215760 & -7.5167457 & 16.06 (0.01) & 15.26 (0.01) & 14.90 (0.02) & 14.48 (0.01) & 17.25 (0.04) & 15.64 (0.03) & 15.10 (0.02) & 14.93 (0.02) & 14.86 (0.01) \\
17 & 41.4300180 & -7.5076398 & 14.74 (0.01) & 14.08 (0.01) & 13.81 (0.02) &              & 14.92 (0.02) & 14.43 (0.02) & 14.18 (0.02) & 13.92 (0.01) & 13.95 (0.01) \\
18 & 41.5925440 & -7.5310037 & 16.06 (0.01) & 13.76 (0.01) & 14.90 (0.02) &              &              &              &              &              &              \\
\hline

\end{tabular}
\\[0.1ex]
\end{scriptsize}
\end{table*}

\begin{table*}
\caption{Positions and photometry of the local sequence reference stars in the 2MASS $JHK$ system.\label{local1}}
\begin{tabular}{llllll} 
\hline 
Star ID & $\alpha_{J2000.0}$ & $\delta_{J2000.0}$ & $J$  & $H$   & $K$   \\ 
               &     (deg)          &    (deg)           & (mag)& (mag) & (mag)  \\ 
\hline 
1      & 41.5216674 & -7.5597940 & 14.82 (0.04) & 14.08 (0.05) & 13.94 (0.05) \\
2      & 41.5496917 & -7.6416869 & 14.32 (0.03) & 13.87 (0.04) & 13.73 (0.05) \\
3      & 41.5474764 & -7.6530580 & 14.71 (0.04) & 14.35 (0.05) & 14.14 (0.06) \\
12     & 41.5528922 & -7.5585795 & 13.63 (0.03) & 13.01 (0.03) & 12.81 (0.03) \\
\hline
\end{tabular} 
\end{table*}

Photometric measurements were carried out with the \texttt{QUBA} pipeline \citep{Valenti2011}, which performs \texttt{DAOPHOT}-based \citep{Stetson1987} point-spread-function (PSF) fitting photometry on the SN and on the selected reference stars. Since SN 2012ec is embedded in a spiral arm of the host galaxy, the background was estimated with a polynomial model. We performed empirical tests for the best background subtraction, and in most cases we found that a $4\mathrm{th}$-order polynomial model of the background gave satisfactory results, due to the high S/N ratio of the SN in these images. Only at the last few epochs was the S/N ratio of the SN too low to prohibit satisfactory removal of the local background. We note, however, that even using the subtraction of a template image would probably not yield a significant improvement, as in these cases the flux of the SN was only few tens of counts above the local background. Photometric uncertainties were automatically estimated by the pipeline using artificial star experiments.

The photometric measurements of the SN in the $BVRI$, $u'g'r'i'z'$ and in the $JHK$ filter systems are reported in Table \ref{phot}.

\begin{table*}
\caption{Optical photometry in the Johnson-Cousins filters, in $u'g'r'i'z'$ bands and NIR photometry calibrated to the 2MASS system, with associated errors in parentheses.\label{phot}}
\centering
\resizebox{\textwidth}{!}{ %
\begin{tabular}{llcccccccccccc}
\hline
Date & $MJD$  & $B$ & $V$  & $R$  & $I$ & $u'$ & $g'$ & $r'$ & $i'$ & $z'$ & $J$ & $H$ & $K$  \\
& & (mag) & (mag) & (mag) & (mag) & (mag) & (mag) & (mag) & (mag) & (mag) & (mag) & (mag) & (mag) \\
\hline
20120814 & 56154.22 & 14.99 (0.02) & 14.81 (0.02) & & & 15.02 (0.04) & & 14.78 (0.02) & 14.91 (0.02) & & & &  \\
20120815 & 51155.22 & 14.99 (0.04) & 14.86 (0.04) & & & 15.09 (0.02) & & 14.81 (0.02) & 14.91 (0.02) & & & &  \\
20120817 & 56157.59 & 15.12 (0.06) & 14.90 (0.06) & 14.74 (0.06) & 14.55 (0.05) & & & & & & & & \\
20120818 & 56158.23 & 15.10 (0.03) & 14.87 (0.03) & & & 15.28 (0.05) & & 14.82 (0.01) &14.94 (0.01) & & & & \\
20120819 & 56158.34 & 15.15 (0.06) & 14.95 (0.05) & 14.73 (0.06) & 14.53 (0.03) & & & 14.84 (0.05) & 14.86 & & & &  \\
20120820 & 56159.31 & 15.29 (0.05) & 14.92 (0.04) & 14.62 (0.05) & 14.53 (0.05) & & 15.05 (0.07) & 14.78 (0.03) & 14.86 (0.06) & & & & \\
20120821 & 56160.30 & 15.18 (0.07) & 14.86 (0.06) & 14.65 (0.03) & (14.61 (0.06) & & 15.13 (0.10) & 14.81 (0.04) & 14.89 (0.04) & 14.92 (0.05) & & & \\
20120826 & 56165.28 & 15.47 (0.05) & 14.93 (0.04) & 14.64 (0.04) & & 16.35 (0.04) & 15.25 (0.08) & 14.80 (0.03) & 14.87 (0.03) & 14.92 (0.03) & 14.24 (0.02) & 14.04 (0.02) & 13.91 (0.02) \\
20120828 & 56168.20 & 15.55 (0.02) & 15.02 (0.02) & & & 16.52(0.06) & & 14.85 (0.02) & 14.93 (0.02) & & & & \\
20120831 & 56171.08 & 15.67 (0.06) & 14.98 (0.06) & & & 16.69 (0.08) & & 14.81 (0.02) & 14.94 (0.02) & & & & \\
20120902 & 56173.09 & 15.76 (0.04) & 14.99 (0.04) & & & 16.98 (0.06) & & 14.85(0.02) & 14.92 (0.02) & & & & \\
20120905 & 56176.13 & 15.76 (0.04) & 15.00 (0.05) & 14.65 (0.06) & 14.62 (0.06) &17.05 (0.07) & & 14.84 (0.03) & 14.86 (0.03) & 14.89 (0.03) & & & \\
20120909 & 56179.34 & 15.90 (0.06) & 15.10 (0.06) & 14.78 (0.04) & 14.45 (0.02) & & 15.54 (0.02) & & 14.92 & & 14.11 (0.03) & 13.89 (0.03) & 13.82 (0.03) \\
20120910 & 56180.92 & 15.95 (0.04) & 15.14 (0.03) & 14.76 (0.01) & 14.51 (0.01) & & & & & & & & \\
20120911 & 56181.59 & 16.05 (0.02) & 15.15 (0.02) & 14.79 (0.03) & 14.53 (0.03) & & & & & & & & \\
20120916 & 56186.20 & 16.06 (0.07) & 15.10 (0.06) & 14.75 (0.04) & 14.49 (0.02) & & 15.51 (0.04) & 14.89 (0.04) & 14.87 (0.02) & 14.85 (0.03) & & & \\
20120920 & 56190.24 &              & 15.12 (0.03) & & 14.42 (0.05) & & & 14.89 (0.03) & 14.85 (0.03) & 14.87 (0.03) & & & \\
20120923 & 56194.87 & 16.15 (0.02) & 15.10 (0.02) & 14 78 (0.01) & 14.45 (0.01) & & & & & & & & \\
20120924 & 56195.23 & & & & & & & & & & 14.08 (0.03) & 13.89 (0.03) & 13.75 (0.03) \\
20120926 & 56196.20 & 16.16 (0.06) & 15.00 (0.05) & 14.72 (0.03) & & 17.96 (0.09) & 15.56 (0.03) & 14.81 (0.03) & 14.81 (0.03) & 14.78 (0.03) & & & \\
20120929 & 56199.29 &              & 15.02 (0.03) & 14.74 (0.03) & 14.36 (0.01) & & & 14.88 (0.02) & 14.81 (0.02) & 14.79 (0.02) & & & \\
20121001 & 56202.01 & 16.19 (0.05) & 15.11 (0.05) & & & 17.98 (0.11) & & 14.89 (0.02) & 14.80 (0.03) & 14.82 (0.02) & & & \\
20121002 & 56202.20 & 16.28 (0.04) & 15.04 (0.04) & 14.74 (0.04) & 14.40 (0.02) & & 15.61 (0.06) & 14.93 (0.05) & 14.84 (0.04) & 14.85 (0.04) & & & \\
20121004 & 56204.21 & 16.33 (0.04) & 15.12 (0.03) & 14.73 (0.03) & 14.43 (0.02) & 18.07 (0.14) & 15.71 (0.04) & 14.90 (0.03) & 14.84 (0.02) & 14.84 (0.02) & & & \\
20121007 & 56208.04 & 16.23 (0.08) & 15.11 (0.08) & & & 18.07 (0.18) & & 14.85 (0.02) & 14.78 (0.02) & 14.81 (0.03) & & & \\
20121010 & 56211.05 & 16.35 (0.04) & 15.16 (0.04) & & & 18.26 (0.10) & 15.68 (0.03) & 14.89 (0.01) & 14.83 (0.01) & 14.85 (0.02) & & & \\
20121012 & 56212.19 & 16.46 (0.06) & 15.22 (0.07) & 14.78 (0.03) & 14.41 (0.02) & & & 14.94 (0.02) & 14.79 (0.03) & 14.87 (0.03) & & & \\
20121016 & 56216.35 & & & & & & & & & & 14.05 (0.03) & 13.80 (0.03) & 13.59 (0.03) \\
20121017 & 56217.15 & 16.50 (0.07) & 15.22 (0.06) & & & & & 14.95 (0.03) & 14.89 (0.03) & 14.77 (0.04) & & & \\
20121019 & 56220.42 & 16.63 (0.05) & 15.28 (0.05) & 14.77 (0.04) & 14.41 (0.03) & & & & & & & & \\
20121020 & 56221.06 & 16.58 (0.03) & 15.36 (0.03) & 14.8 (0.1) & 14.43 (0.02) &18.68 (0.08) & & 14.96 (0.02) & 14.86 (0.02) & 14.92 (0.02) & & & \\
20121022 & 56223.52 & & & & & & & 15.00 (0.02) & 14.85 (0.03) & & & & \\
20121024 & 56226.45 & & & & & & 15.99 (0.06) & 15.01 (0.02) &14.91 (0.03) & & & & \\
20121101 & 56232.13 & 16.79 (0.09) & 15.47 (0.08) & 14.95 (0.04) & 14.59 (0.02) & & 16.00 (0.06) & 15.15 (0.03) & & 15.05 (0.03) & & & \\
20121106 & 56237.12 &              & 15.63 (0.03) & 15.04 (0.03) & 14.67 (0.03) & & 16.2 (0.1) & 15.26 (0.02) & 15.15 (0.02) & 15.16 (0.02) & 14.35 ( 0.06) & 14.12 (0.06) & 14.04 (0.04) \\
20121111 & 56242.13 & 17.1 (0.1) & 15.85 (0.08) & 15.26 (0.03) & 14.85 (0.03) & & 16.38 (0.06) & 15.39 (0.02) & 15.29 (0.02) & 15.26 (0.03) & & & \\
20121114 & 56245.20 & & & & & & & & & &14.58 (0.03) & 14.34 (0.03) & 14.28 (0.03) \\
20121115 & 56246.96 & 17.4 (0.1) & 16.09 (0.10) & & & & & 15.63 (0.02) & 15.47 (0.02) & 15.43 (0.02) & & & \\
20121117 & 56248.14 &              & 16.26 (0.04) & 15.57 (0.05) & 15.15 (0.04) & & & & & & & & \\
20121119 & 56250.19 & 17.82 (0.09) & 16.49 (0.08) & 15.85 (0.05) & 15.37 (0.05) & & 17.24 (0.05) & 16.00 (0.02) & 15.89 (0.02) & 15.74 (0.03) & & & \\
20121122 & 56253.08 & 17.95 (0.10) & 17.16 (0.10) & 16.36 (0.07) & 15.63 (0.05) & & 17.45 (0.10) & 16.32 (0.03) & 16.29 (0.03) & 16.12 (0.12) & & & \\
20121204 & 56266.14 & & & & & & & & & & 15.73 (0.06) & 15.27 (0.08) & 15.40 (0.06) \\
20121205 & 56266.93 & 18.5 (0.2) & 17.3 (0.2) & & & & & 16.65 (0.07) & 16.51 (0.07) & 16.40 (0.06) & & & \\
20121207 & 56268.94 & 18.60 (0.15) & 17.40 (0.15) & & & & & 16.80(0.1) & 16.6 (0.1) & 16.50 (0.06) & & & \\
20121209 & 56270.95 & 18.70 (0.13) & 17.50 (0.13) & & & & & 16.9 (0.1) & 16.7 (0.1) & 16.6 (0.1) & & & \\
20121216 & 56277.99 & & & & & & & 17.0 (0.1) & 16.9 (0.1) & 16.8 (0.1) & & & \\
20121220 & 56282.94 & 18.8 (0.2) & 17.7 (0.2) & 16.9 (0.2) & & & & & & & & & \\
20121221 & 56283.10 & & & & & & & & & & 15.83 (0.03) & 15.41 (0.03) & 15.47 (0.03) \\
20121228 & 56290.00 & 19 (0.2) & 17.9 (0.2) & & & & & 17.1 (0.1) & 17.1 (0.1) & 16.9 (0.1) & & & \\
20130110 & 56302.81 & 19.15 (0.30) & 18.0 (0.3) & & & & & 17.2 (0.1) & 17.2 (0.1) & 17.0 (0.1) & & & \\
20130112 & 56305.66 &              & 18.0 (0.3) & 17.15 (0.30) & 16.75 (0.30) & & & & & & & & \\
\hline
\end{tabular}}
\\[1.4ex]
\end{table*}

\subsection{Data analysis}\label{photoanalysis}

\begin{figure*}
  \centering
  \includegraphics[scale=0.4]{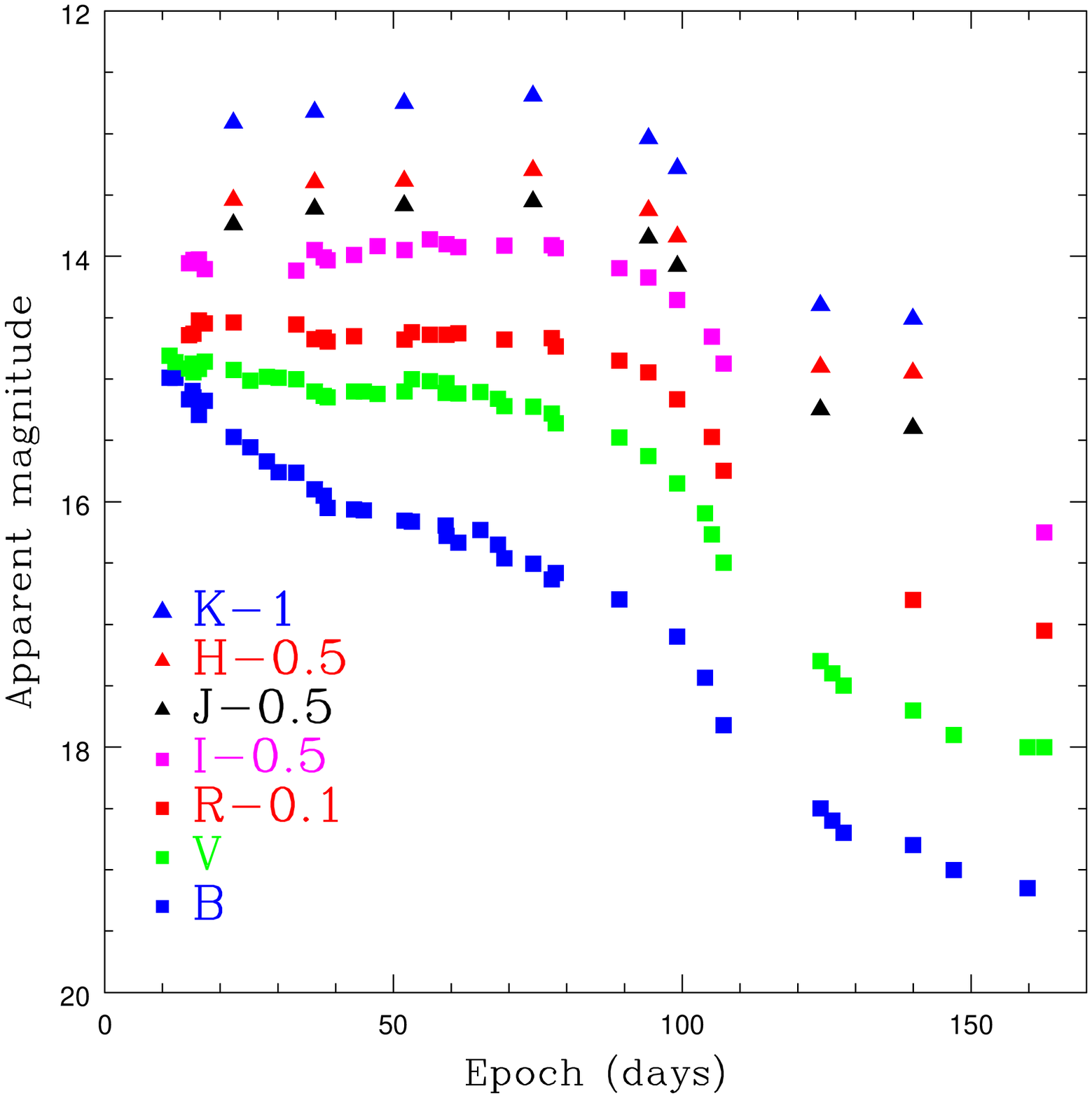}
  \includegraphics[scale=0.4]{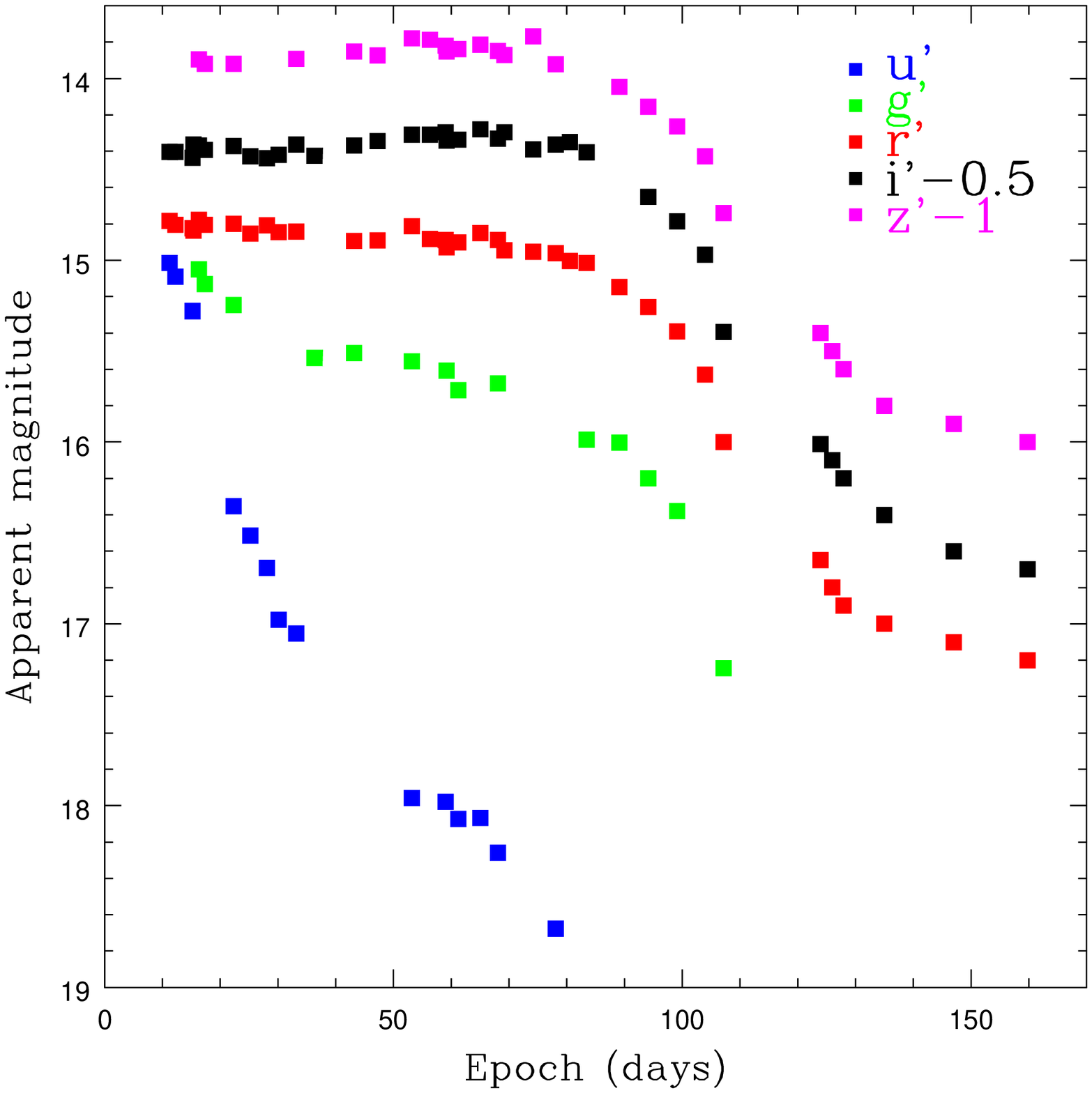}
  \caption{Left panel: photometric evolution of SN2012ec in the Johnson-Cousins $BVRI$ and $JHK$ filters. Right panel: photometric evolution of SN2012ec in the $u'g'r'i'z'$ filters. A shift has been applied for clarity.}
  \label{LC_opt}
\end{figure*}

The photometric evolution of SN 2012ec in the $BVRI$, $JHK$ and in the $u'g'r'i'z'$ filter systems is shown in Fig. \ref{LC_opt}.

SN 2012ec was already on the plateau in the $V, R, I, r',i'$ and $z'$ bands by $+13$ days. The average absolute magnitude, in the different bands, during the plateau phase was $M_{V}=-16.54$ mag, $M_{R}=-16.75$ mag, $M_{I}=-16.96$ mag, $M_{r'}=-16.80$ mag, $M_{i'}=-16.93$ mag and $M_{z'}=-17.08$ mag.
Using the definition for the plateau duration proposed by \citet{Olivares2010}, where the end of the plateau occurs at the knee of the light curve, we found that the plateau of SN 2012ec lasted almost $90$ days in $R, I, r', i',z'$ and almost $80$ days in $V$. This is shorter than the usual duration of the plateau of standard Type II-P SNe (e.g. SN 2004et, $~ 100$ days, \citealt{Maguire2010}; SN 2012aw, $~ 100$ days, \citealt{DallOra2014}; see also \citealt{Arcavi2012}). SN 2012ec began to fall from the plateau at $\sim +90$, while the photospheric phase from the observed spectroscopic evolution (see Sect. \ref{spec}) lasted until $\sim 160$ days.  The decline in the light curve of SN 2012ec, from the plateau to the radioactive decay tail, lasted $\sim 30$ days, decreasing $ ~1.5$ mag in $r',i',V$ bands, $~1$ mag in the $I$ bands and $~1.3$ mag in the $z'$ band.  A list of the main characteristics of the light curve, for each filter, is reported in Table \ref{LCdata}.

\begin{table*}
\setlength{\tabcolsep}{5pt}
\caption{Epochs and apparent magnitudes of the light curve during the plateau in the $VRIr'i'z'$ bands.\label{LCdata}}
\setlength{\tabcolsep}{2.5pt}
\begin{footnotesize}
\begin{tabular}{lrrrrrrrrr}
\hline         
 & V  & R & I & $r'$ & $i'$ & $z'$ & J & H & K  \\
 & mag & mag & mag & mag & mag & mag & mag & mag & mag \\
 \hline
m$_\mathrm{plat}^{a}$ & 15.10 (0.02) & 14.78 (0.01) & 14.45 (0.01) & 14.89 (0.03) & 14.85 (0.03) & 14.87 (0.03) & 14.08 (0.03) & 13.89 (0.03) & 13.75 (0.03) \\
M$_\mathrm{plat}^{a}$ & -16.54 (0.17) & -16.75 (0.17) & -16.96 (0.17) & -16.80 (0.18) & -16.93 (0.18) & -17.08 (0.18) & -17.24 (0.18) & -17.38 (0.18) & -17.49 (0.18) \\
\hline 

\end{tabular}
\\[1.5ex]
$^a$ Plateau phase refers to 59 days after the explosion at $MJD=56202.0$
\end{footnotesize}
\end{table*}

The NIR light curve exhibits a plateau of duration $\sim 90-100$ days, which subsequently drops over a period of $40$ days by $~1.3$ mag in the $J$ band, $1.1$ mag in the $H$ band and $1.2$ mag in the $K$ band. This behaviour is similar to that observed for other Type II-P SNe (see for example, SN 2012A, \citealt{Tomasella2013}; SN 2012aw \citealt{DallOra2014}).

\begin{figure}
  \centering
  \includegraphics[scale=0.4]{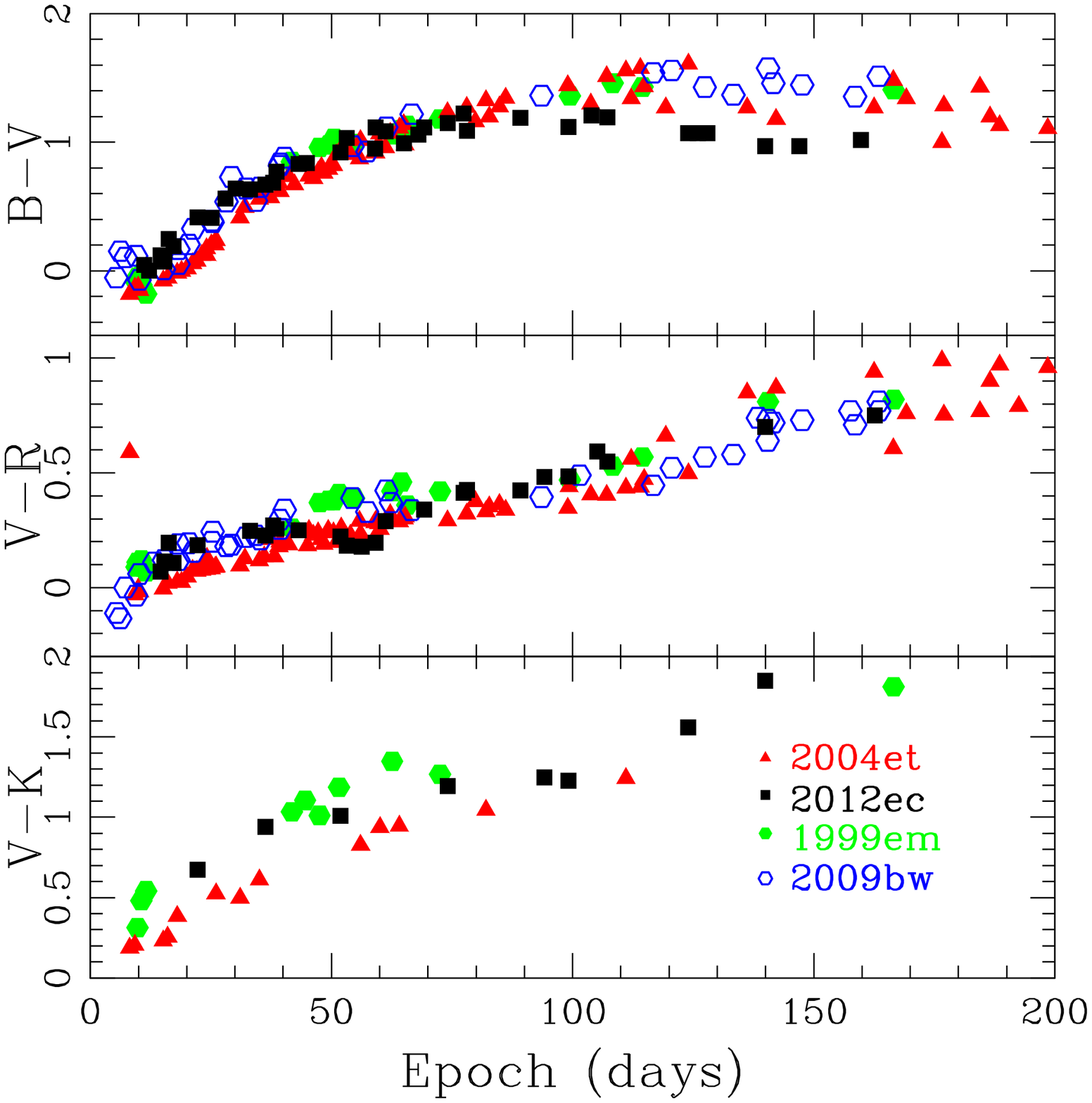}
  \caption{Colour evolution of SN 2012ec compared to other type II-P SNe.}
  \label{color_all}
\end{figure}

The evolution of the $B-V$, $V-R$ and $V-K$ colours of SN 2012ec are shown in Fig. \ref{color_all}. The $B-V$ colour becomes progressively redder over the first $50$ days, rising from $B-V \sim 0$ to $\sim 1\mathrm{mag}$, before reaching a constant value by $\sim 160\mathrm{d}$.  The $V-K$ colour starts from $0.7$ mag and increases slowly to $\sim 1$ mag at $\sim 100$ days, before increasing further  from $\sim 1$ to $\sim 1.9$ mag in the period $100-130$ days.
The colour evolution of SN 2012ec is similar to those of other type II-P SNe (e.g. SN 2004et, \citealt{Maguire2010}; SN 1999em, \citealt{Elmhamdi2003}; SN 2009bw, \citealt{Inserra2012}).
The trends in the colour evolution are similar to those observed by \citealt[][see their Fig. 10]{Faran2014} for a sample of 23 type II-P SNe.

\subsection{Bolometric light curve and $^{56}Ni$ mass}\label{bolometric}

\begin{figure}
  \centering
  \includegraphics[scale=0.4]{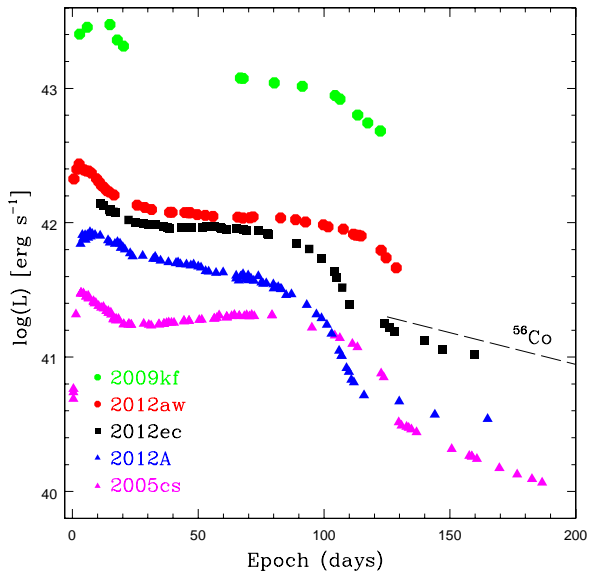}
  \caption{Pseudo-bolometric light curve of SN2012ec, along with those of other type II-P SNe. The pseudo-bolometric light curve accounts for the UBVRIJHK contributions for SN 2012A, UBgVrRiIzJHK for SN 2012aw, griz for SN 2009kf and UBVRIJHK for SN 2005cs.}
  \label{bolom_all}
\end{figure}

A pseudo-bolometric light curve was calculated by integrating over the optical and NIR photometry.
The $u'Bg'Vr'Ri'Iz'JHK$ apparent magnitudes have been converted into monochromatic fluxes at the effective wavelength for each filter, and then corrected for extinction (Sect. \ref{hostgalsec}). The resulting SED was integrated over the entire wavelength range, assuming zero flux at the limits. The estimation of the flux was performed at only those phases for which V band observations were available.  If photometry for other bands was not available, the magnitudes were estimated at these phases by interpolating the values from photometry acquired on adjacent nights.
The final integrated fluxes were converted to luminosity through application of the adopted distance modulus.  The pseudo-bolometric light curve of SN 2012ec is shown in Fig. \ref{bolom_all}.  The luminosity at the first epoch for which the calculation could be conducted ($14$ days) was $L= \mathrm{1.4 \times 10^{42} \: erg \: s^{-1}}$; this can be considered a lower limit for the bolometric luminosity.  The SN luminosity reaches the plateau by day $20$ ($L= \mathrm{0.9 \times 10^{42} \: erg \: s^{-1}}$), which then begins to significantly decrease at $\sim 90$ days to the tail at day $~130$, with a luminosity of $L= \mathrm{0.1 \times 10^{42} \: erg \: s^{-1}}$. 

A comparison of the pseudo-bolometric light curve of SN~2012ec with other Type II-P SNe demonstrates a similar behaviour (e.g. SN 2012A, \citealt{Tomasella2013}; SN 2012aw, \citealt{DallOra2014}; SN 2009kf, \citealt{Botticella2010}; and SN 2005cs, \citealt{Pastorello2009}). From the pseudo-bolometric light curve of SN 2012ec, it is evident that its luminosity on the plateau is lower than observed for SNe 2012aw and SN 2009kf and that plateau duration is shorter than the more luminous SNe.
SN 2012ec is more luminous than SN 2012A and SN 2005cs but has a behaviour more similar to SN 2012A. They have comparable plateau, even if the one of SN 2012A is a bit shorter.
Instead SN 2005cs shows a different evolution of the light curve compared to SN 2012ec, especially the fall from the plateau that is longer for SN 2005cs.

We estimated the $\mathrm{^{56}Ni}$ mass synthesised during the explosion, by comparing the luminosity of SN 2012ec with that of SN 1987A at similar late epochs.
Assuming a similar $\gamma$-ray deposition fraction, the mass of $\mathrm{^{56}Ni}$ was calculated using the relation of \citet{Bouchet1991}:

\begin{equation}
\label{eq:nichel}
M(^{56}Ni)_{12ec} = M(^{56}Ni)_{87A} \times \frac{L_{12ec}}{L_{87A}} (\msun)
\end{equation}

For the $\mathrm{^{56}Ni}$ mass of SN 1987A we adopted the weighted mean of the values reported by \citet{Arnett1989} and \citet{Bouchet1991}, and for the bolometric luminosity we adopted the value of \citet{Bouchet1991} (see also \citealt{Suntzeff1988}).  For SN 2012ec we calculated $M(^{56}Ni)_{12ec}= 0.040 \pm 0.015 \: \msun$, which is an average of the estimates made at $138$, $146$ and $158$ days (the reported uncertainty is the dispersion of the values computed at each epoch).
 The slope of the light curve in the last epochs of the dataset is $0.01 \pm 0.02 \mathrm{\: mag \: day^{-1}}$, in agreement with the $^{56}Co$ rate of decay.
  The data from the nebular phase are published in a companion paper (\citealt{Jerkstrand14b}, submitted). \citet{Jerkstrand14b} estimate the nickel mass from photometry at 187 and 202 days, finding a value of $0.03 \pm 0.01 \: \msun$, which is in good agreement with our estimate.

\begin{figure}
  \centering
  \includegraphics[scale=0.75]{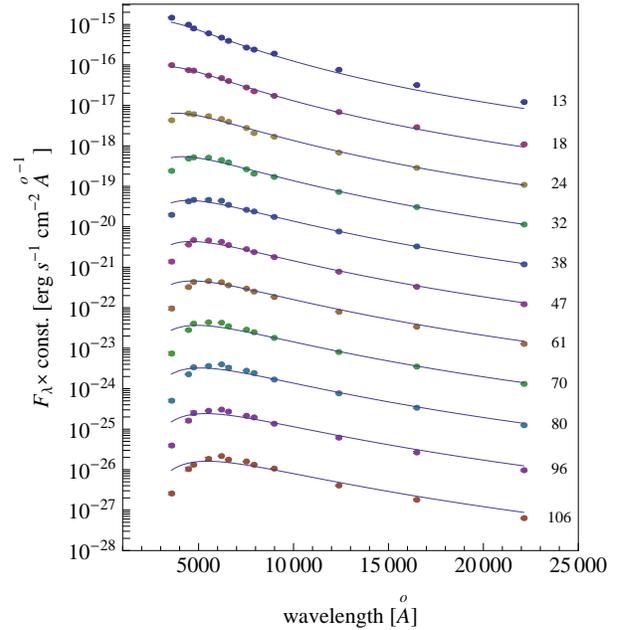}
  \caption{The temporal evolution of the SED of SN 2012ec. Circles represent the fluxes at the central wavelengths of each filter. Solid lines represent blackbody continuum fits. Fluxes are corrected for distance and extinction.}
  \label{SED}
\end{figure}

The evolution of the SED of SN 2012ec, based on optical and NIR photometry, is shown in Fig. \ref{SED}. The observations covered the wavelength range $4000-23000$ \AA.
We evaluated the evolution of the SED and calculated blackbody continuum fits at each epoch. At $13$ days, the best fit gives a blackbody temperature of $9600 \pm 800$ K, which decreases to $5300 \pm 400$ K by day $106$.  At early time, the fits were conducted using all available photometric observations.  At later epochs, the bluest photometric observations were excluded from the fits as metal line blanketing, particularly due to Fe II and Ti II, at these wavelengths caused significant departures from the ideal black body assumption \citep{Dessart2005}.  The u band data was excluded from the fits for data after $\mathrm{20d}$ and, in addition, the B and g bands were excluded from fits for data after $\mathrm{50d}$.

From the blackbody fit it was possible to evaluate the time evolution of the photospheric temperature of SN 2012ec.
  The temperature drops rapidly in the first $30$ days from $9600 \pm 800$ K to $7000 \pm 500$ K,  before decreasing slowly from $6500 \pm 500$ K to $5000 \pm 400$ K.
  The values of the temperature estimated from the blackbody fits to the photometric data are in good agreement with those derived from fits of the continuum in the observed spectra (within the uncertainties) from $\mathrm{+30d}$. During the first $30$ days the spectroscopic temperature varies from $11000 \pm 900$ K to $8000 \pm 700$ K, decreasing to $6200 \pm 500$ at $~50$ days before reaching $5000 \pm 500$ K in the last epochs.
  The slightly higher temperatures estimated from the spectra are due to the limited spectroscopic wavelength range ($4000-9000$ \AA) used for the continuum fits, compared to the wavelength range covered by the available photometric data.
We compared the estimated temperatures with those of SNe 2009bw \citep{Inserra2012} and 1999em \citep{Elmhamdi2003}.
SN 2012ec is cooler at earlier phases, compared to  SN 2009bw which had an initial temperature of $\sim 12000$ K and SN 1999em which had a temperature of $\sim 14300$ K.  At later pahases, the temperatures of all three SNe converge to $\sim 5000$ K.

\section{Spectroscopic evolution}

\subsection{Data sample and reduction}\label{Specsec}

As a PESSTO follow-up target, SN 2012ec was scheduled for a dense spectroscopic monitoring campaign at the ESO NTT at La Silla, Chile. Ten epochs of optical spectroscopy were acquired with  EFOSC2 and ten epochs of NIR spectroscopy were acquired with SOFI.
The optical dataset was supplemented with spectra from the following facilities: the $2.3$m telescope of the Siding Spring Observatory (SSO, New South Wales, Australia) equipped with the Wide Field Spectrograph WiFeS ($2$ epochs), the $2.5$m Nordic Optical Telescope (NOT, Canary Islands, Spain) equipped with the Andalucia Faint Object Spectrograph and Camera (ALFOSC) ($1$ epoch), the 1.82m Copernico Telescope (Asiago, Italy) equipped with AFOSC ($3$ epochs), the William Herschel Telescope (WHT, Canary Islands, Spain) equipped with the Intermediate dispersion Spectrograph and Imaging system (ISIS) ($1$ epoch), the 1.22m Galileo Telescope (Asiago, Italy) equipped with the Boller $\&$ Chivens spectrograph (B$\&$C) ($2$ epochs).
The spectroscopic observations cover $29$ epochs from day $8$ to day $161$.
Details of the spectroscopic observations and the characteristics of the instruments used are listed in Table \ref{logspec}.

\begin{table*}
\caption{Summary of instrumental sets-up used for the spectroscopic follow-up campaign.}\label{logspec}
\begin{footnotesize}
\begin{tabular}{lclllcll}
\hline
Telescope & Instrument & Grism & Range & Resolution & \# of epochs  \\
 &  & & [ \AA\ ] & [ \AA\ ] & \\
\hline
NTT (3.58m)    & EFOSC2 & Gr11, Gr16  & 3350-10000 & 12 & 10 \\
NTT (3.58m)    & SOFI   & GB          & 9400-14000 & 20 & 7 \\
NTT (3.58m)    & SOFI   & GB, GR      & 14000-25000 & 20 & 3 \\
CAO (1.82m)    & AFOSC  & Gr4         & 3500-8200  & 24 &  3 \\
Pennar (1.22m) & B$\&$C & Gr300       & 3400-7800  & 10 &  2 \\
NOT (2.56m)    & ALFOSC & Gr4         & 3400-9000  & 14 &  1 \\
WHT (4.2m)     & ISIS   & R300B+R158R & 3500-10000 &  5 &  1 \\
ANU (2.3m)     & WiFeS  & B+R         & 3300-9000  &  2 &  2 \\
\hline
\end{tabular}
\\[1.5ex]
NTT = New Technology Telescope with the optical camera ESO Faint Object Spectrograph and Camera EFOSC2 and with the Near-Unfrared Camera Son of ISAAC (SOFI); CAO = the Copernico telescope at Asiago Observatory with the Asiago Faint Object Spectrograph and Camera (AFOSC); Pennar = Galileo telescope at Asiago Observatory with the Boller $\&$ Chivens spectrograph; NOT = Nordic Optical Telescope with the Andalucia Faint Object Spectrograph and Camera (ALFOSC); WHT = William Herschel Telescope with the Intermediate dispersion Spectrograph and Imaging System (ISIS); ANU = Australian National University telescope with the Wide-Field Spectrograph (WiFeS).

\end{footnotesize}
\end{table*}

Spectra were pre-reduced (trimmed, overscan, bias and flat-field corrected) using the PESSTO pipeline (\citealt{Smartt2014}, submitted), based on the standard IRAF tasks \footnote{Fast reduction data are available on WISeREP \citep{Yaron2012} and full reduced data can be accessed from the ESO Phase 3 archive, all details on www.pessto.org}. The wavelength calibration was performed using comparison spectra of arc lamps acquired with the same instrumental configuration as the SN observations. The science observations were flux calibrated with respect to observations of spectrophotometric standard stars. Further corrections for atmospheric extinction were applied using tabulated extinction coefficients for each telescope site (in the pipeline archive).

The quality of the flux calibration was checked by comparison of synthetic $BV$ and $r$ photometry derived from the spectra, using the IRAF task \texttt{CALCPHOT}, with the observed photometry at comparable epochs.  Calibrated spectra were finally dereddened for the total line-of-sight extinction and then corrected for the heliocentric velocity of the host galaxy (see Table \ref{galprop}).

\subsection{Data analysis} 
\label{spec}

\begin{figure*}
  \includegraphics[scale=.9, angle=0]{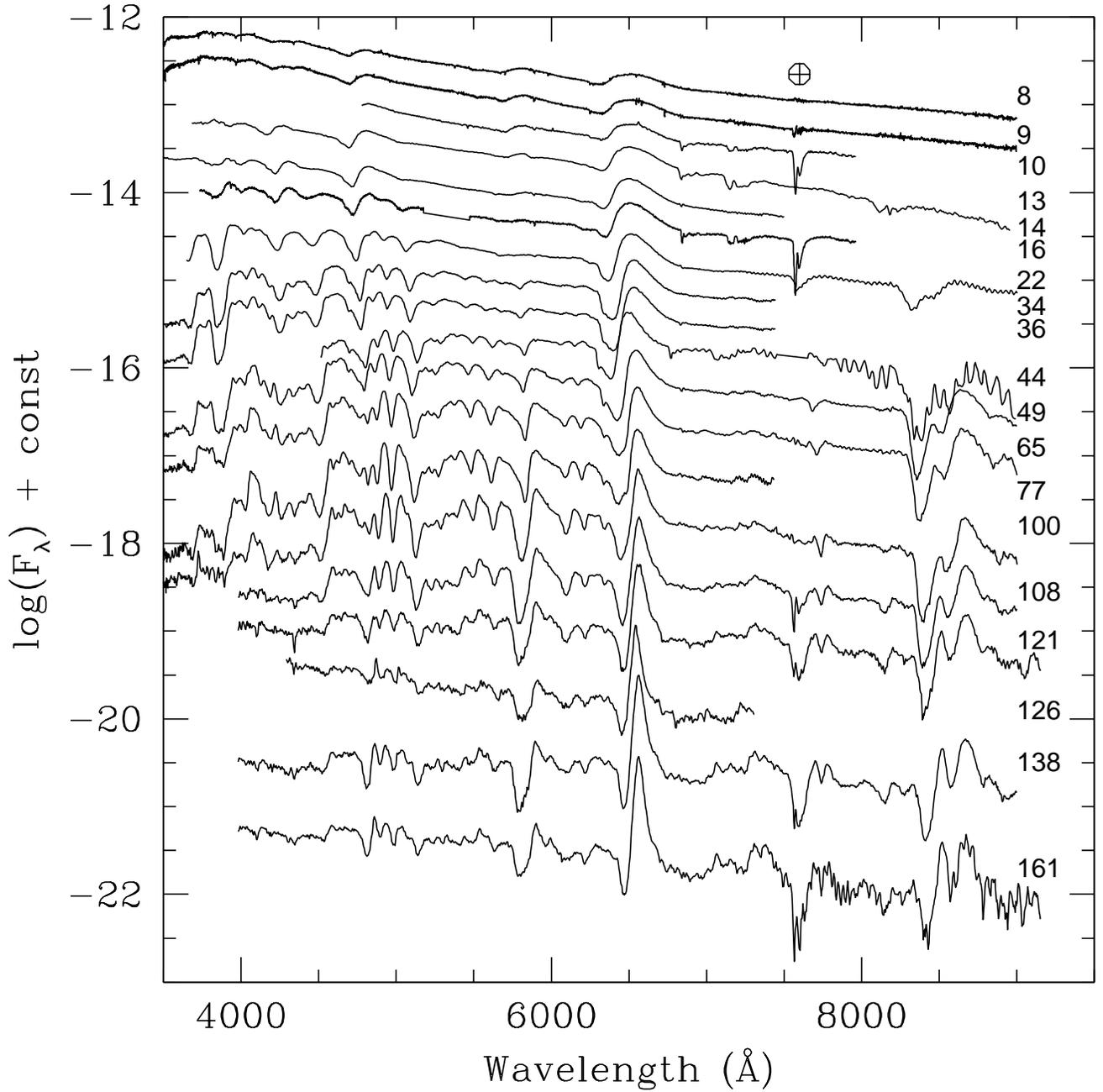}
  \caption{The optical spectroscopic evolution of SN2012ec during the photosperic phase, from $+8$ to $+161$ days.}
  \label{spec_evol}
\end{figure*}

\begin{figure*}
  \centering
  \includegraphics[scale=.5, angle=0]{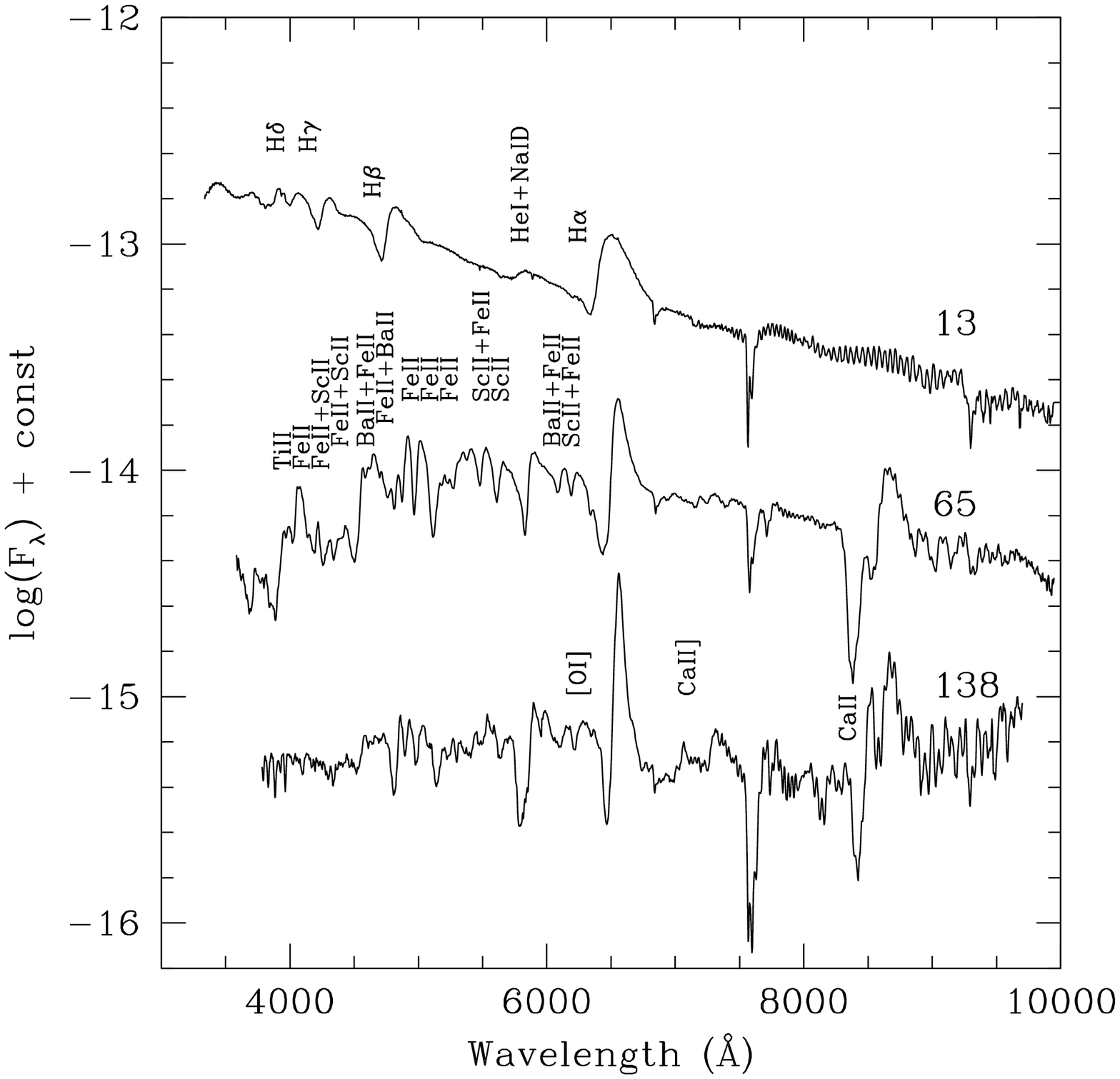}
  \includegraphics[scale=.5, angle=0]{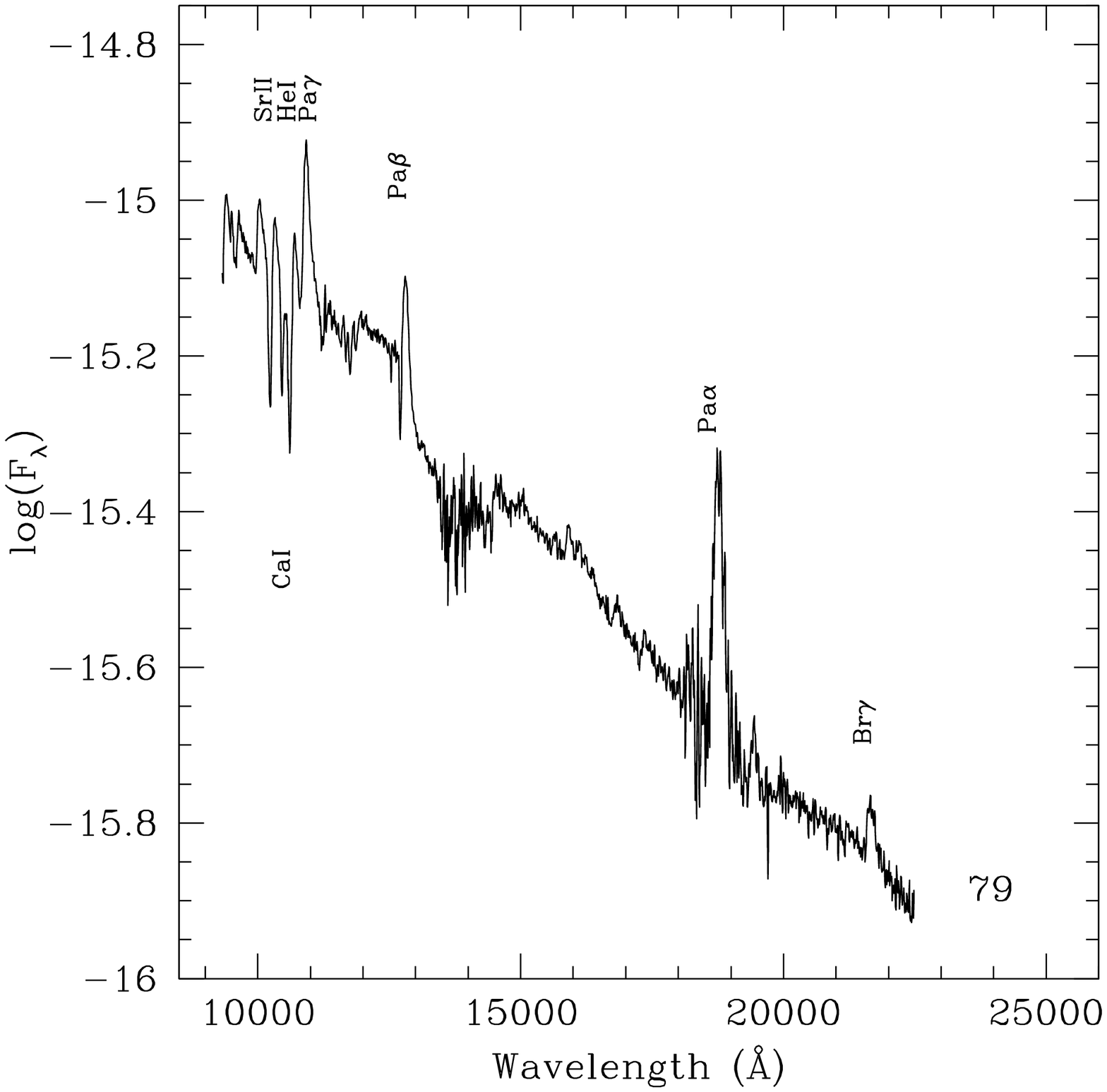}
  \caption{Identifications of line features observed in optical (at three characteristic epochs; top panel) and NIR spectra (bottom panel) of SN 2012ec.}
  \label{spec_id}
\end{figure*}

The time evolution of the optical spectrum of SN 2012ec, obtained from 8 to 161 days, is shown in Fig. \ref{spec_evol} and corresponding line identifications are presented in Fig. \ref{spec_id}.

\begin{figure}
  \centering
  \includegraphics[scale=0.45]{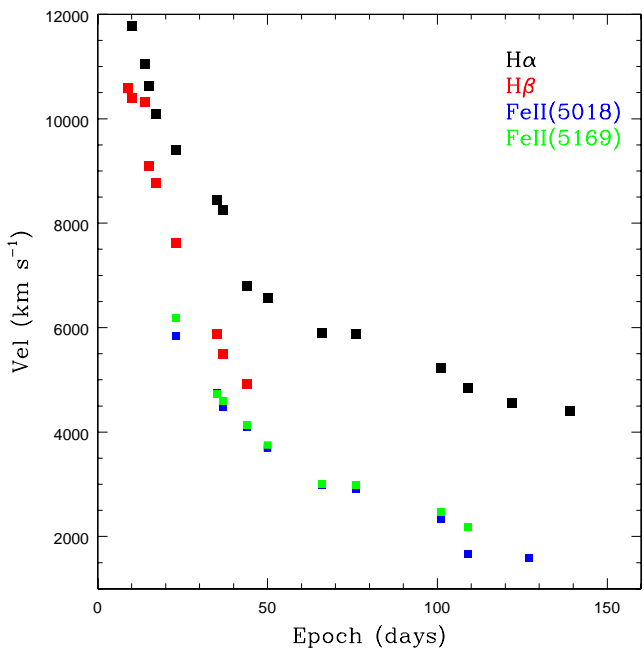}
  \caption{Ejecta velocity evolution, estimated from the H$_\alpha$, H$_\beta$, Fe II(5018 \AA) and Fe II(5169 \AA) lines.}
  \label{vel}
\end{figure}

Fig. \ref{vel} shows the evolution of the velocities of H$_\alpha$, H$_\beta$, Fe II(5018 \AA) and Fe II(5169 \AA) for SN 2012ec. A list of line velocities is presented in Table \ref{tabvelocities}.

Spectra at early phases show a blue continuum, broad Balmer lines and He I at $5876$ \AA.
Lines show the typical P-Cygni profile, from which we estimate expansion velocities from the measurement of the position of the minimum of the absorption component.
At early times, the estimated velocities are $\mathrm{12200 \pm 150 \: km \: s^{-1}}$ for $H_{\alpha}$, $\mathrm{11000 \pm 150 \: km \: s^{-1}}$ for $H_{\beta}$ and $\mathrm{10500 \pm 150 \: km \: s^{-1}}$ for He~I.
A blackbody fit to the continuum of these spectra, in the range $4000-9500$ \AA\, yielded a temperature $11900 \pm 900 \: K$.

Spectra from day 21 to day 44 show, in addition to the Balmer lines, some iron-group elements like Fe II (4629 \AA), Fe II (5018 \AA), Fe II (5169 \AA) and Sc II (6246 \AA). There is also a feature at $8200$ \AA\ due to the Ca II infrared triplet.
The $H_{\alpha}$ velocity decreases to $10000 \pm 120 \: km \: s^{-1}$, $H_{\beta}$ to $9000 \pm 120 \: km \: s^{-1}$, while the velocities for the Fe II(5018 \AA) and Fe II(5169 \AA) were measured to be $\sim 6000 \pm 100 \: km \: s^{-1}$.  The temperatures derived from blackbody fits to the continuum show a decrease from $8000 \pm 500 \: K$ to $6000 \pm 300 \: K$.

Spectra from day 49 to day 138 show the appearance of lines due to other heavy elements, such as Ba II(5981 \AA), Ba II(6142 \AA), Ti II(4100 \AA), and numerous blends of Fe II lines, while  the absorption feature of NaID is no longer visible. At early times, the NaID  feature is clearly visible as an absorption on the continuum, but at later times it is blended with complex broad features.
At these phases the velocities decrease for all elements: the velocity of $H_{\alpha}$ decreases to $5000 \pm 90 \: km \: s^{-1}$ and Fe II (5018 \AA) and Fe II (5169 \AA) decrease to $2000 \pm 120 \: km \: s^{-1}$.
The presence of the iron-group line blends prevents the detection of $H_{\beta}$.  A fit to the continuum yields a temperature of $5000 \pm 400 \: K$.

At late times, the spectrum at 161 days shows forbidden [O I] lines (6300, 6364 \AA) and the semi-forbidden Ca II] doublet (7291, 7394  \AA).

The ejecta velocities of SN 2012ec have been compared with those measured for other Type II-P SNe: SN 2012A, SN 2012aw, SN 2004et and SN 1999em (see Table \ref{vel_SN}).
At early phases, the $H_{\alpha}$ velocity is lower than that estimated for SN 2012aw ($\mathrm{\sim 14000 \: km \: s^{-1}}$;, \citealt{DallOra2014}), but higher that the one estimated for SN 2012A ($\sim10200, \: km \: s^{-1}$) \citep{Tomasella2013}, and comparable with the one of SN 1999em ($\sim 12000, \: km \: s^{-1}$) \citep{Elmhamdi2003}.
At later phases ($40$ days), the Fe II (5169 \AA) velocities are higher than those estimated for SN 2012A ($\sim 3500 \: km \: s^{-1} $), comparable with those of SN 2004et ($\sim 4000 \: km \: s^{-1} $) and SN 1999em ($\sim 4200 \: km \: s^{-1} $), but they are still lower than that of SN 2012aw ($\sim 5500 \: km \: s^{-1}$).
In summary, the ejecta velocities measured for SN 2012ec velocities are similar to those measured for SNe 1999em and 2004et, but are consistently lower than for SN 2012aw and higher than for SN 2012A.
We also point out that the evolution of the Fe II(5169), $H_{\alpha}$ and $H_{\beta}$ velocities of SN 2012ec are in excellent agreement with the trends shown in Figure 16 of \citet{Faran2014}, based on a sample of 23 well-studied II-P SNe.

\begin{table*}
\caption{Expansion velocity of SN 2012ec at selected epochs, compared to other Type II-P SNe.\label{vel_SN}}
\setlength{\tabcolsep}{2.5pt}
\begin{tabular}{lccccc}
\hline
& 2012aw & 2012ec & 1999em & 2004et & 2012A \\
\hline
$H_{\alpha}$ ($\sim 10$ d) & 14000 & 12200 & 12000 & & 10200 \\
Fe II ($\sim 40$ d) & 5500 & 4100 & 4200 & 4000 & 3500 \\
Fe II ($\sim 100$ d) & 3000 & 2400 & 2000 & 2000 & 2000 \\
\hline
\end{tabular}
\end{table*}

\begin{table*}
\caption{Measured expansion velocities (from the minima of P-Cygni absorption) for SN 2012ec. 
Estimated uncertaintes are in parentheses}\label{tabvelocities}
\begin{scriptsize}
\begin{tabular}{rrrrrrrrr}
\hline
Date	 & MJD    & $Epoch^{a}$  & $H_{\alpha}$   & $H_{\beta}$ & $Fe II(5018)$  & $Fe II(5169)$ & $ScII(5533)$ & $Ca II(8520)$  \\
 &  &  (d) & $km \: s^{-1}$ & $km \: s^{-1}$ & $km \: s^{-1}$ & $km \: s^{-1}$ & $km \: s^{-1}$ & $km \: s^{-1}$ \\
\hline
20120812 & 56152 & 8     &  12200 (150)     & 10600	(150)   &         	  &       	    &       	    &             \\
20120813 & 56153 & 9	 &  11800	(130)	& 10400	(150)	&         	  &       	    &            	&             \\
20120817 & 56157 & 13    &  11000	(160)	& 10300	(130)	&         	  &       	    &           	&             \\
20120818 & 56158 & 14    &  10600	(120)	&  9100	(120)	&         	  &       	    &           	&             \\
20120820 & 56160 & 16    &  10100	(120)	&  8800	(110)	&            	&       	    &           	&             \\
20120826 & 56166 & 22    &   9400	(100)	&  7600	(120)	& 5800	(100)	& 6200	(100)	&           	&             \\
20120907 & 56178 & 34    &   8400	(110)	&  5900	(110)	& 4700	(100)	& 4700	(120)	& 5000	(120)	&             \\
20120909 & 56180 & 36    & 	 8300	(110)	&  5500	(130)	& 4500	(110)	& 4600	(100)	& 4600	(130)	&             \\
20120916 & 56187 & 43    & 	 6800	(120)	&  4900	(110)	& 4100	(110)	& 4100	(130)	&           	&             \\
20120922 & 56193 & 49    & 	 6600	(110)	&           	& 3700	(100)	& 3700	(100)	& 3800	(140)	& 5600	(120) \\
20121008 & 56209 & 56    & 	 5900	(110)	&           	& 3000	(100)	& 3000	(140)	& 3100	(100)	& 4900	(140) \\
20121017 & 56219 & 75    & 	 5800	(170)	&           	& 2900	(110)	& 2900	(150)	&           	&             \\
20121112 & 56244 & 100   & 	 5230	(120)	&           	& 2300	(120)	& 2400	(100)	& 2100	(130)	& 4100	(100) \\
20121122 & 56252 & 108	 & 	 4800	(100)	&         	    & 				& 2200	(100)	& 2000	(150)	& 3700	(150) \\
20121203 & 56265 & 121	 & 	 4500	(100)	&           	& 2000	(110)	&           	&           	& 3600	(130) \\
20121212 & 56270 & 126	 &           	    &           	& 1600	(100)	&           	&           	&             \\
20121220 & 56282 & 138	 &   4400	(100)	&           	&         	  &           	&           	& 3500	(140) \\

\hline 

\end{tabular}
\\[1.4ex]
$a$ = epoch from the explosion.
\end{scriptsize}
\end{table*}

\begin{figure}
  \centering
  \includegraphics[scale=.4, angle=0]{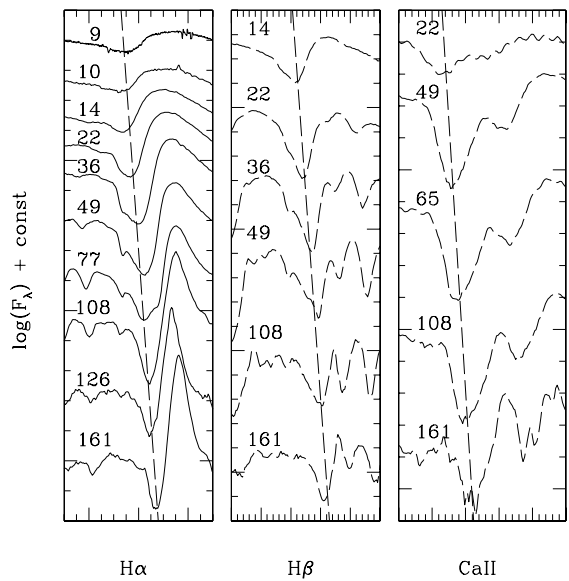}
  \caption{Time evolution of $H_{\alpha}$, $H_{\beta}$ and Ca II NIR triplet for SN 2012ec.}
  \label{lines}
\end{figure}

A close-up showing the time evolution of the $H_{\alpha}$, $H_{\beta}$ and Ca II line profiles for SN 2012ec is shown in Fig. \ref{lines}.

\begin{figure*}
  \centering
  \includegraphics[scale=.75, angle=0]{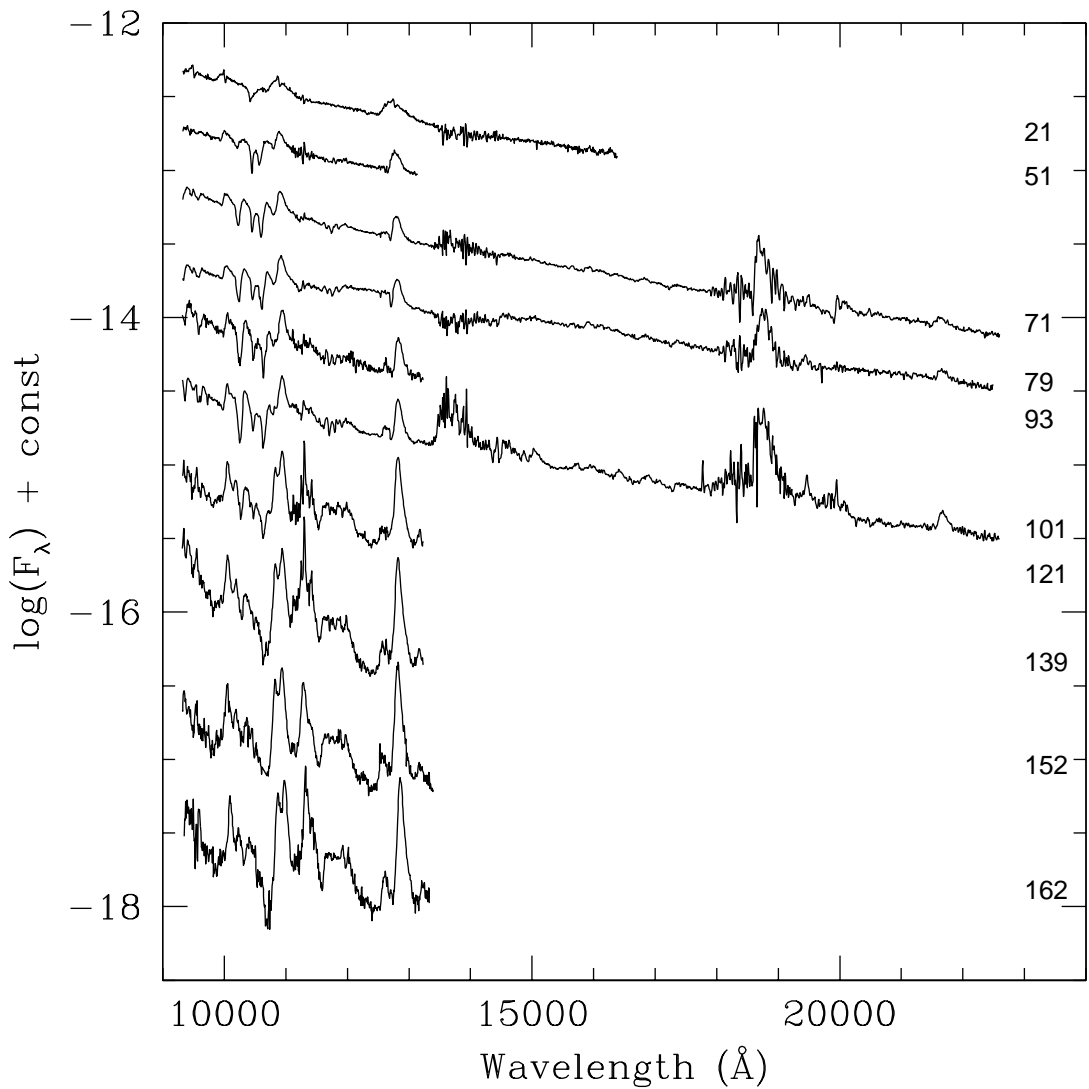}
  \caption{NIR spectroscopic evolution of SN 2012ec. Individual spectra have been shifted in flux for clarity. Numbers on the right indicate the epochs from explosion.}
  \label{spec_nir}
\end{figure*}

The NIR spectra cover the period from day $21$ to  day $161$ (Fig. \ref{spec_nir}). The H I Paschen lines are clearly visible at all epochs. Starting from day $68$ we identify also He I and Ca I lines and $Br_{\gamma}$.
The elements identified in the NIR spectra (Fig. \ref{spec_id}) are typical of Type II-P SNe, in particular the spectra at $71$ and $79$ days are similar to the NIR spectrum of SN 2012A  at 72 days \citep{Tomasella2013}.

\section{Hydrodynamic modeling}\label{modelling}

To constrain the main physical properties of the progenitor and the energetics of the explosion, we performed hydrodynamical modelling of SN 2012ec. Among the most important parameters we need to constrain are the ejected mass, the radius of the progenitor, the explosion energy and the ejected $\mathrm{^{56}Ni}$ mass (\citealt{Zampieri2003}; \citealt{Kasen2009}). These were found by comparing the observed bolometric luminosity, the evolution of line velocities and continuum temperature at the photosphere with the corresponding simulated quantities (\citealt{Zampieri2003}; \citealt{Pumo2010}). The comparison procedure consists of performing a simultaneous $\chi^{2}$ fit of all the relevant observables against those predicted by the model calculations.  This approach was successfully adopted for other CC-SNe (e.g. SN 2007od, \citealt{Inserra2011}; SN 2009bw, \citealt{Inserra2012}; SN 2009E, \citealt{Pastorello2012}; SN 2012A, \citealt{Tomasella2013}; and SN 2012aw, \citealt{DallOra2014}).

The hydrodynamical modelling of the explosion was performed with two different codes: a semi-analytic code \citep{Zampieri2003}, that solves the energy balance equation for a constant density envelope which expands homologously; and a radiation-hydrodynamics code \citep{Pumo2011}, that can simulate the full radiative-hydrodynamical evolution of the ejected material. The latter code solves the hydrodynamic equations of a self-gravitating, relativistic fluid interacting with radiation, and incorporates an accurate treatment of radiative transfer and of the evolution of the ejected material, considering both the gravitational effect of the compact remnant and the heating effects related to the decays of radioactive isotopes synthesized during the CC SN explosion. The first code is used to investigate the more likely parameter space and provide a robust, first estimate of the best fitting model. A more detailed and time-consuming search is then performed with the radiation-hydrodynamics code. This modeling is appropriate only if the emission from the CC SN is dominated by freely expanding ejecta. Clearly, interaction with the circumstellar medium (CSM) can affect the early evolution of the light curve in a way not presently predicted by the models.

\begin{figure}
  \centering
  \includegraphics[width=0.45\textwidth ]{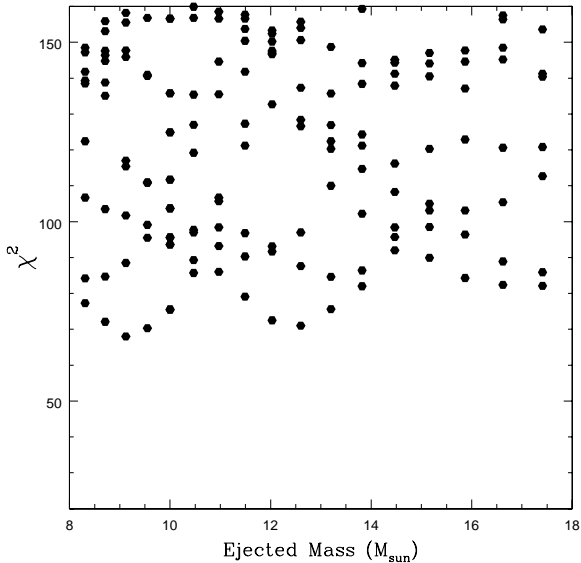}
  \caption{$\chi^{2}$ distribution of the fit of the semi-analytical model to the observed quantities, as a function of the estimated ejected mass.}
  \label{mod-chi}
\end{figure}

\begin{figure}
  \centering
  \includegraphics[width=0.45\textwidth ]{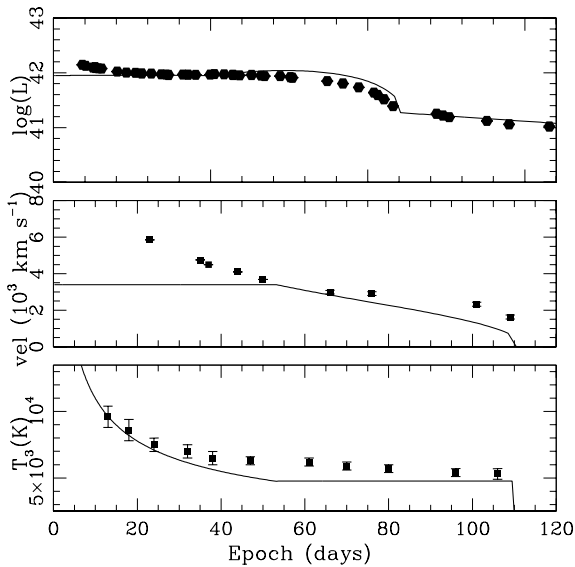}
 \caption{Time evolution of the main observables of SN 2012ec (filled dots), compared to the ``high-mass'' best fit model (solid line). The top panel shows the fit of the bolometric light curve; the middle panel shows the the fit of the Fe II velocity and the bottom panel shows the fit of the continuum temperature.}
  \label{mod-fit}
\end{figure}

An extended grid of semi-analytic models was computed, covering a wide range in mass. The $\chi^{2}$ distribution of the models as a function of ejected mass is shown in Fig. \ref{mod-chi} and shows two comparable minima, one at $\sim 9.1 \: \msun$, the other at $\sim 12.6 \: \msun$. 
The best fit model corresponding to the first minumum ($9.1 \pm 0.8 \: \msun$) has an initial radius of $\sim 2.3 \times 10^{13} \pm 0.7 \: cm$ ($330 \pm 100 \: R_{\odot}$), a total explosion energy of $\sim 0.7 \pm 0.2 \: foe$ and an ejected $^{56}$Ni mass of $\sim 0.035 \: \msun$.       
The model corresponding to the second minumum has an initial radius of $1.6 \pm 0.5 \times 10^{13} \: cm$ ($230 \pm 70 \: R_{\odot}$), a total explosion energy of $1.2 \pm 0.4 \: foe$, and an ejected $^{56}$Ni mass of $\sim 0.035 \: \msun$.
In light of the results of the progenitor detection in pre-explosion observations, we only consider the ``high-mass'' minimum further.
The best fit model corresponding to the second minimum is shown in Fig. \ref{mod-fit} and appears to be in good agreement with all the observables.

\section{Homogeneous comparison with the two well-studied II-P SNe 2012A and 2012aw}

In this section, we present a detailed comparison of SN 2012ec with two well studied Type II-P SNe: 2012A and 2012aw.  In all three cases, a progenitor was detected in pre-explosion images and sufficient photometric and spectroscopic observations were available to permit a homogenous analysis of the properties of the SNe using the same hydrodynamical code.
SN 2012ec was discovered 9 days after the explosion, while the other SNe were discovered much sooner after explosion (see Table \ref{comp_SNe}).  SN 2012aw was discovered in M95 at a distance modulus $\mu=29.96 \pm 0.04$ mag and with a total reddening of $E(B-V)=0.086$ mag; while SN 2012A was discovered in NGC 3239 at $\mu=29.96 \pm 0.15$ and $E(B-V)=0.037$ mag.

The estimates of the initial masses of the progenitors, through direct detection of the precursor, were: $M_{12aw}= 14-26 \; \msun$ \citep{Fraser2012}, $M_{12ec}$ in the range $14-22 \; \msun$ \citep{Maund2013} and $M_{12A}=8-15 \; \msun$ \citep{Tomasella2013}.  In a separate analysis of the pre-explosion observations of SN~2012aw, \citealt{Van2012} reported an initial mass of $15-20 \: \msun$. A major uncertainty in estimating the progenitor mass is degeneracy between temperature and reddening. \citealt{Kochanek2012} showed that a different treatment of the extincion results in a luminosity of $log(L/L_{\odot})= 4.8-5.0$, corresponding to a progenitor main sequence mass of $13-16 \: \msun$ \citep{Jerkstrand2014}, which is in agreement with the nebular spectral modelling and the amount of oxygen produced by SN 2012aw.

\begin{figure}
  \centering
  \includegraphics[scale=0.4]{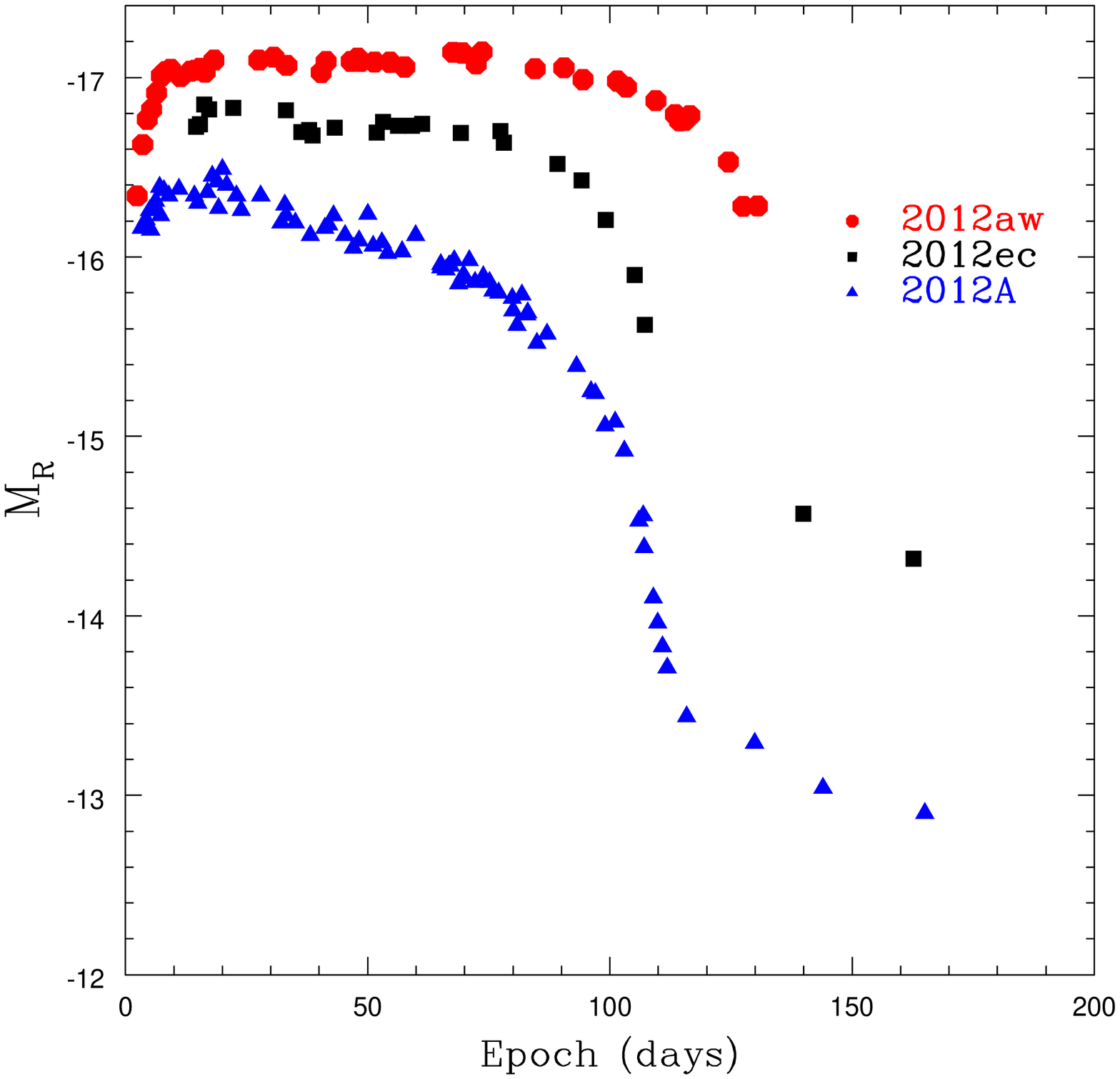}
  \includegraphics[scale=0.4]{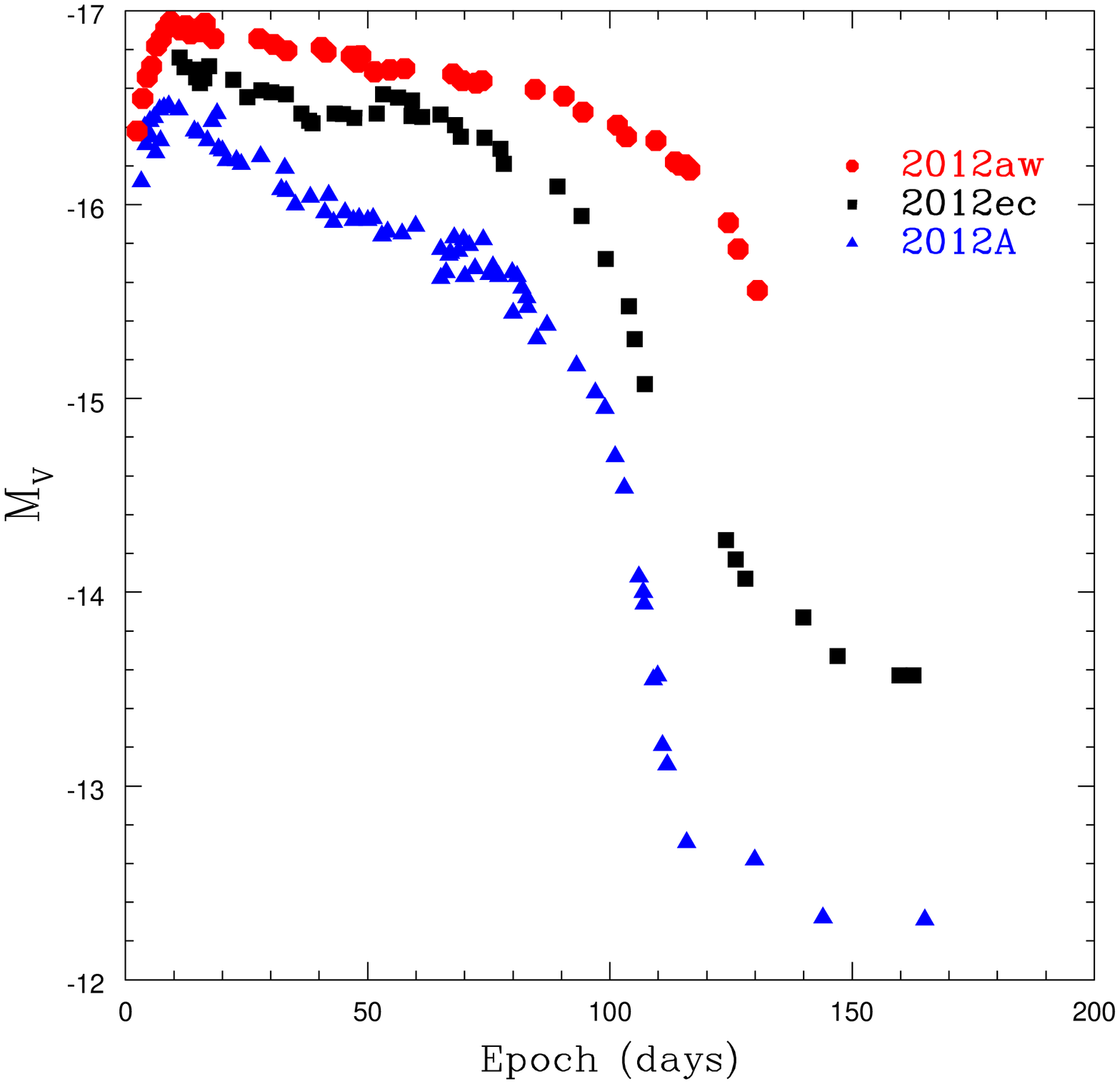}
  \caption{Comparison of the light curves in the $R$ (top panel) and $V$ (bottom panel) bands of SN 2012ec, with SN 2012aw and SN 2012A.}
  \label{R}
\end{figure}

Fig. \ref{R} shows the photometric evolution of the absolute magnitudes in the R and V bands of SN 2012ec, SN 2012aw and SN 2012A.
We note that SN 2012ec is intermediate between the more luminous SN 2012aw and the fainter SN 2012A. The duration of the plateau and the post-plateau decline is longer in SN 2012aw and shorter and steeper in SN 2012A. Again, SN 2012ec shows an intermediate behaviour, with quite a short plateau and a slower post-plateau drop.
The absolute magnitude in the R band for these SNe, on the plateau ($\sim$ 60 days), were $M_{R}(12aw)= -17.1$ mag, $M_{R}(12ec)= -16.7$ mag and $M_{R}(12A)= -16.2$ mag.

\begin{figure}
  \centering
  \includegraphics[scale=0.4]{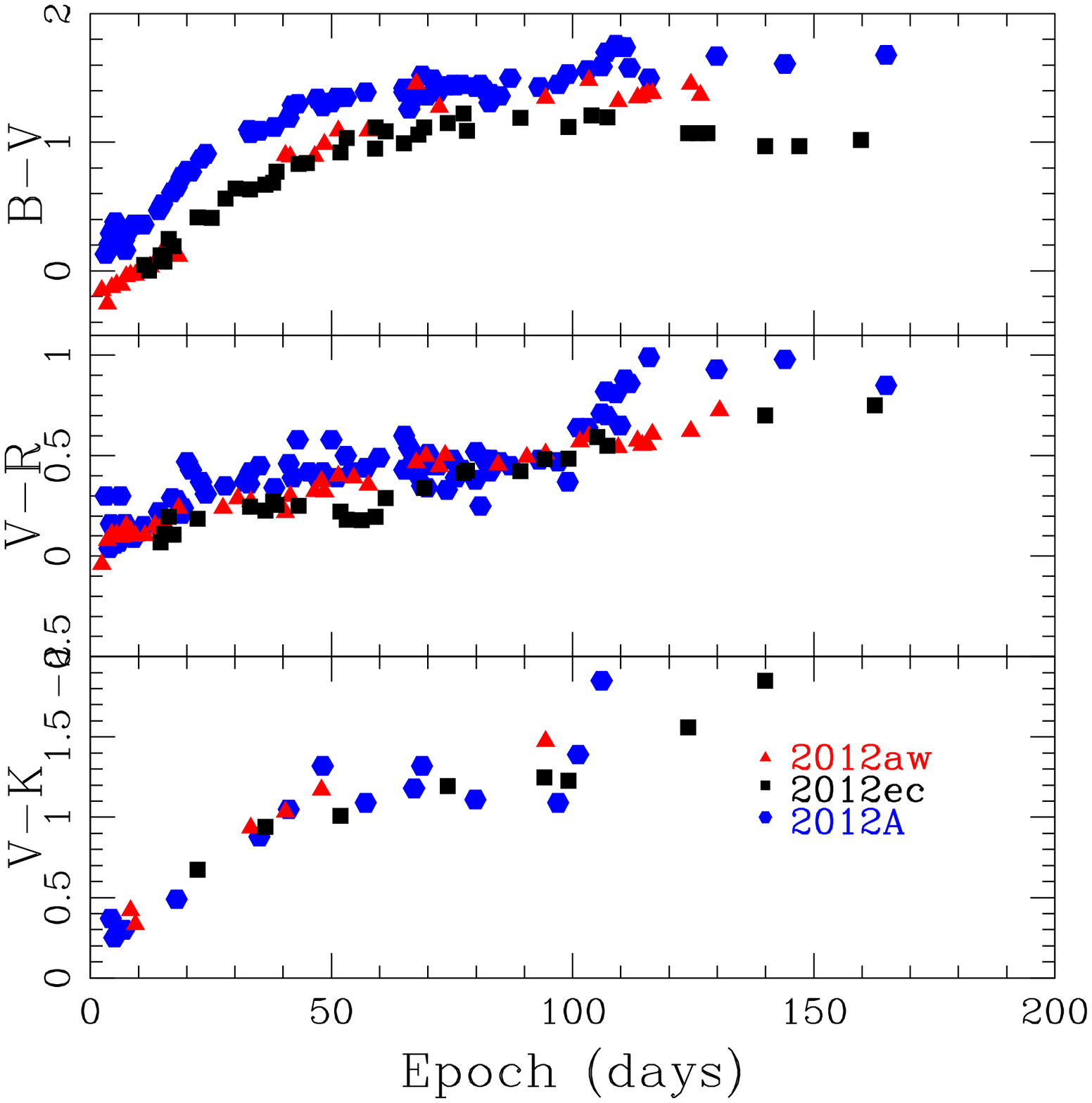}
  \caption{Comparison of the colour evolution of SN 2012ec, in the $B-V$ (top panel), $V-R$ (middle panel), and $V-K$ (bottom panel), with SN 2012aw and SN 2012A.}
  \label{color}
\end{figure}

A comparison of the colours evolution of SN 2012ec with SN 2012aw and SN 2012A is shown in Fig. \ref{color}. The colour of each SN has been corrected for reddening for a proper comparison.
The colour evolution of SN 2012ec has already been discussed in Sect. \ref{photoanalysis}.
From Fig. \ref{color}, we can see that the colour evolution of SN 2012ec is similar to that of the other two SNe.

\begin{figure}
  \centering
  \includegraphics[scale=0.4]{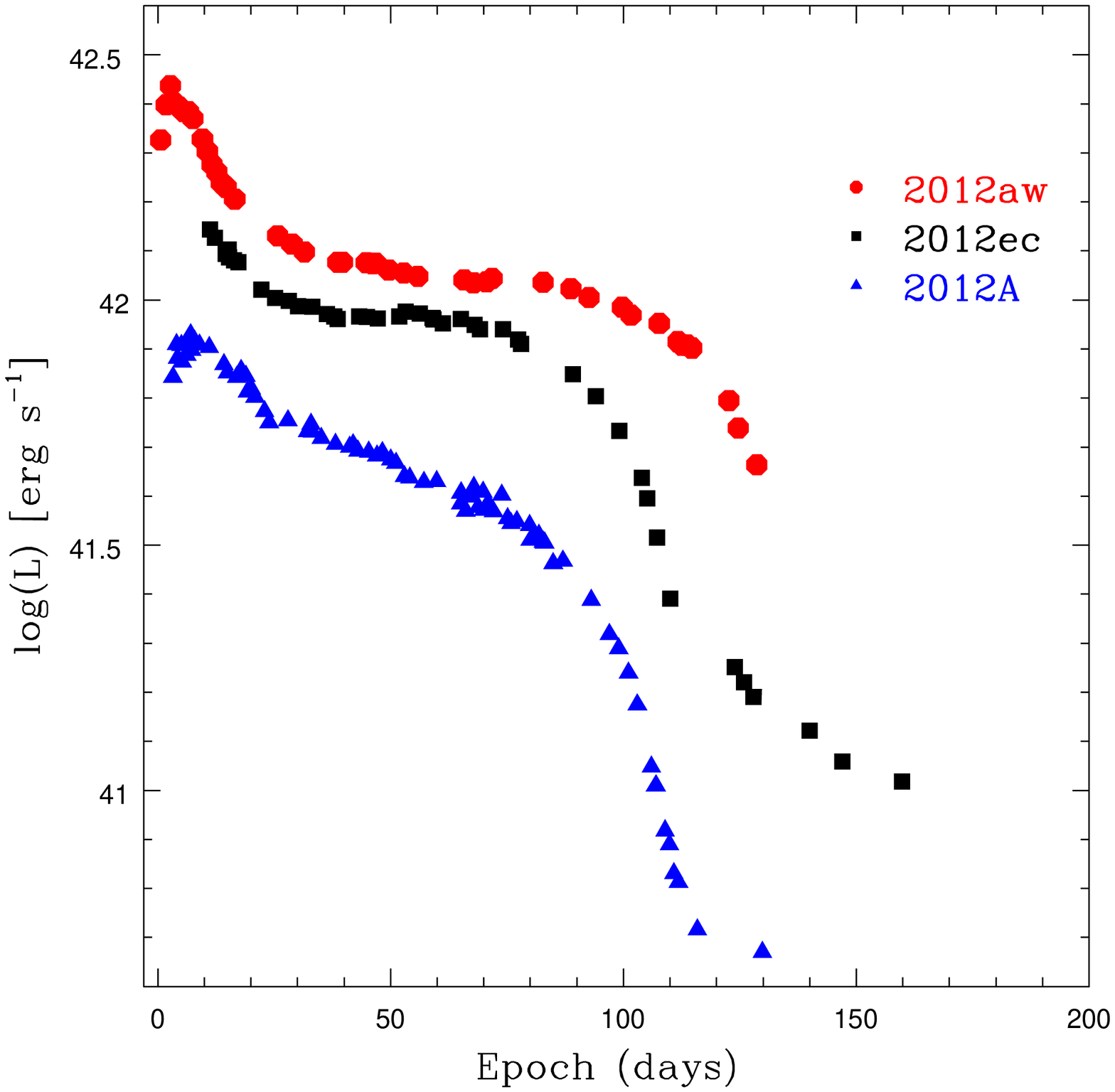}
  \caption{Pseudo-bolometric light curve of SN 2012ec, compared to SN 2012aw and SN 2012A.}
  \label{bolom_comp}
\end{figure}

Fig. \ref{bolom_comp} shows a comparison of the bolometric light curves of SNe 2012ec, 2012A and 2012aw, where SN 2012ec is of intermediate  luminosity between the other two SNe. In particular, during the plateau phase, SN 2012ec is more luminous than SN 2012A and exhibits a longer plateau.  Conversely, SN 2012aw is clearly of higher luminosity than SN 2012ec throughout the entirety of the photospheric phase and has a longer plateau of $\sim100$ days \citep{DallOra2014}.

From the comparison of the $^{56}Ni$ masses estimated for the three SNe, we may note a sequence in the values: $M(^{56}Ni)_{12aw}= 0.056 \pm 0.013 \: \msun$, $M(^{56}Ni)_{12ec}= 0.040 \pm 0.015 \: \msun$ and $M(^{56}Ni)_{12A}= 0.011 \pm 0.004 \: \msun$.

\begin{figure*}
  \centering
  \includegraphics[scale=.7, angle=0]{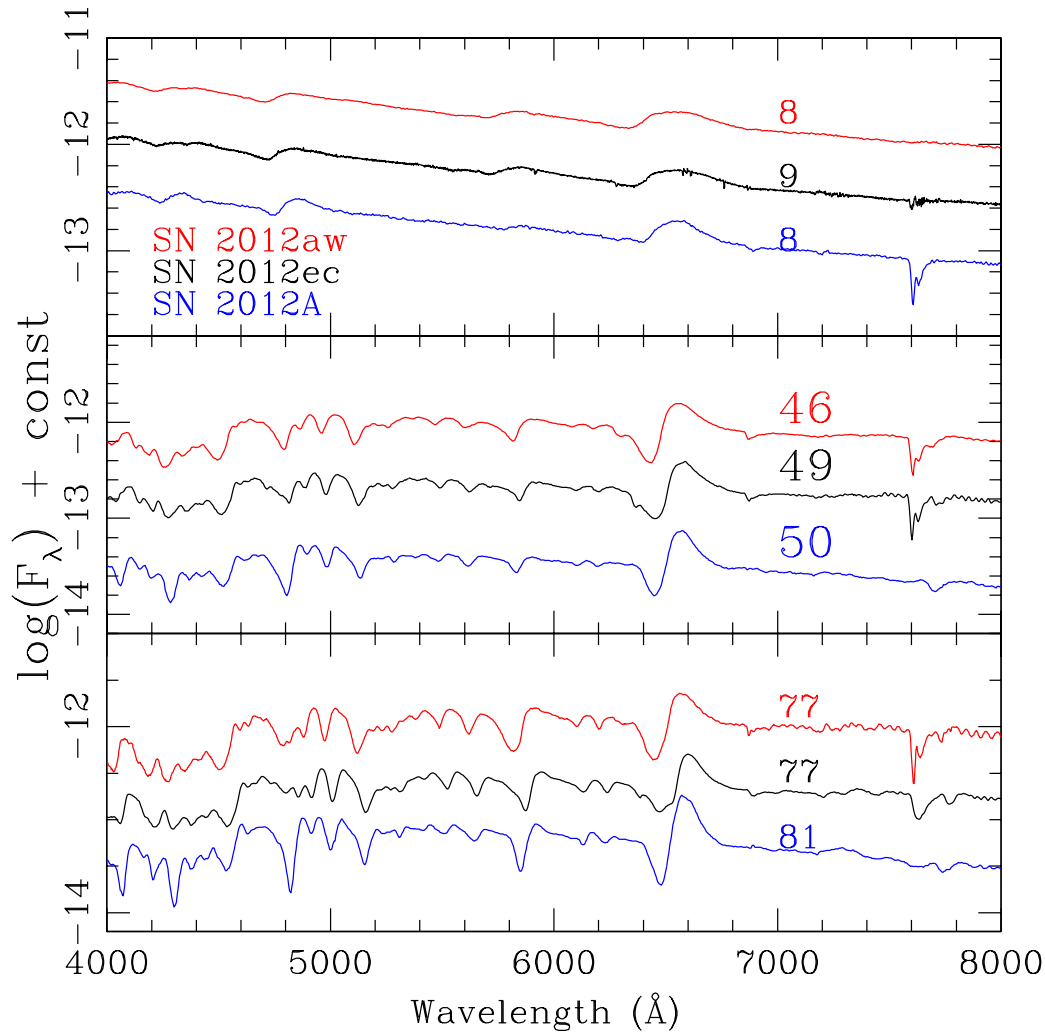}
  \caption{Comparison of the spectra of SN 2012ec, SN 2012aw and SN 2012A at three different epochs, i.e. at early times, during the plateau phase and at the end of the plateau.}
  \label{spec_comp}
\end{figure*}

In Fig. \ref{spec_comp} we show a comparison of the spectra of SN 2012ec with those of SN 2012aw and SN 2012A at three different epochs, highlighting the spectroscopic similarities between the three SNe at all epochs.

\begin{figure}
  \centering
  \includegraphics[scale=0.4]{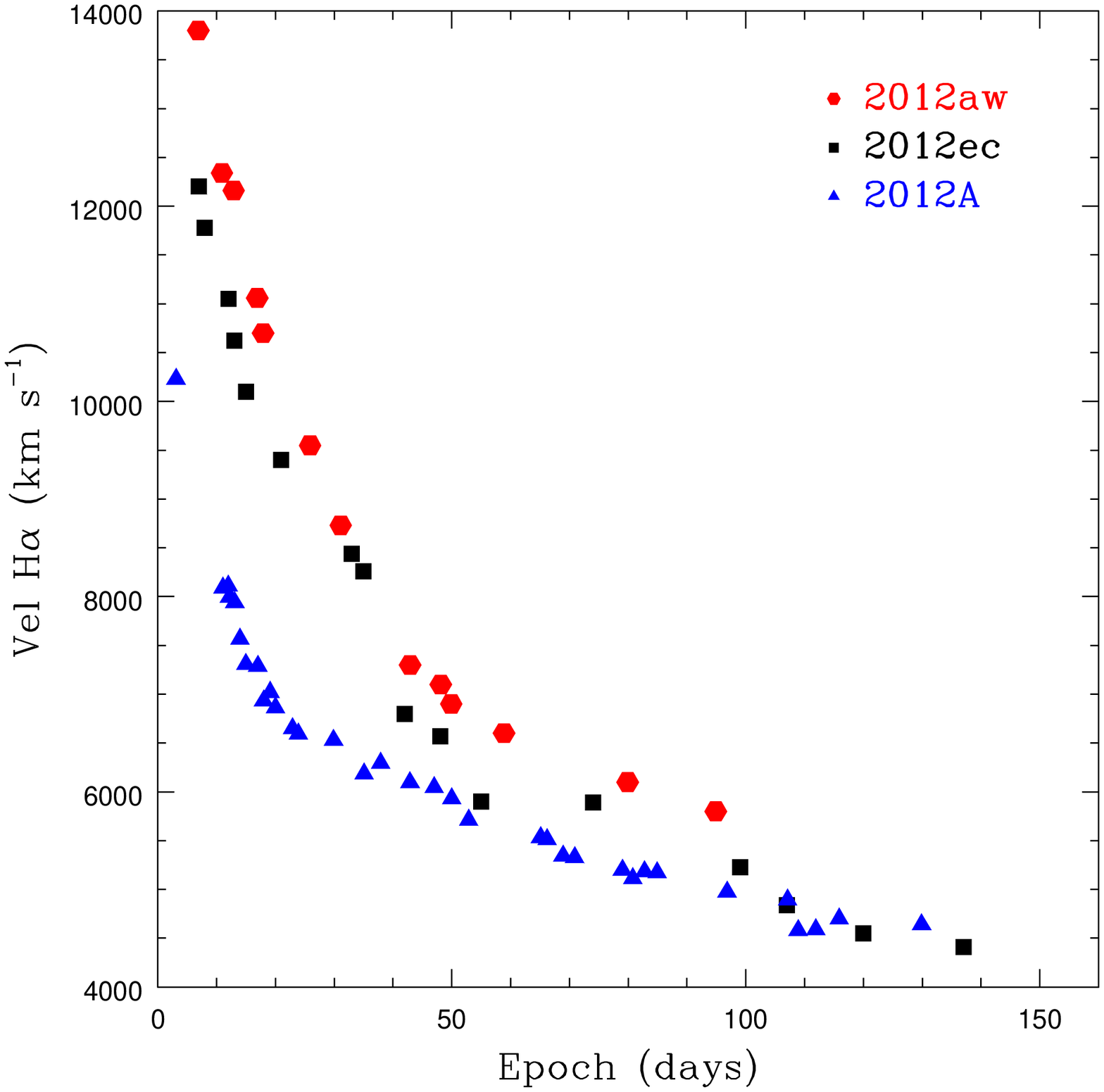}
	\includegraphics[scale=0.4]{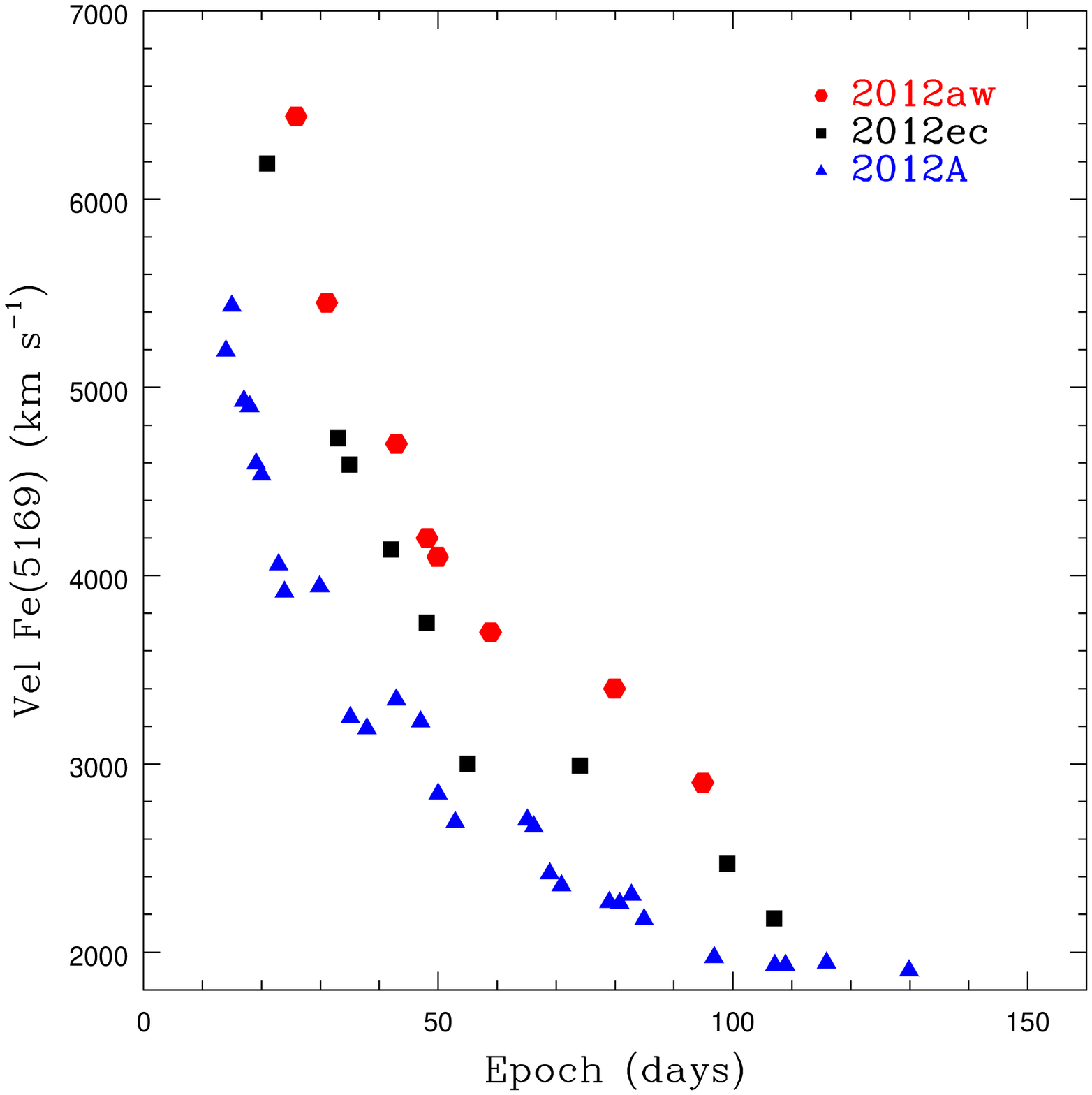}
  \caption{Comparison of the ejecta velocities of SN 2012ec, SN 2012A and SN 2012aw, measured from the H$_\alpha$ (top panel) and  Fe II(5169 \AA) lines (bottom panel).}
  \label{vel_H}
\end{figure}

We also compared the ejecta velocities measured from H$_\alpha$, and Fe II(5169 \AA) for SN 2012ec with the velocities measured for other type II-P SNe (see Fig. \ref{vel_H}). SN 2012aw has an initial $H_{\alpha}$ velocity $\sim 14000 \: km \: s^{-1}$, higher than measured for SN 2012ec ($\sim 12200 \: km \: s^{-1}$) and for SN 2012A ($\sim 10200 \: km \: s^{-1}$). After 100 days, the velocity of $H_{\alpha}$ decreases to $\sim 6000 \: km \: s^{-1}$ for SN 2012aw, which is still higher than measured for SN 2012ec ($\sim 5000 \: km \: s^{-1}$) and for SN 2012A ($\sim 5000 \: km \: s^{-1}$).  The initial Fe II(5169 \AA) of SN 2012aw is $\sim 6500 \: km \: s^{-1}$, still higher than those of SN 2012ec ($\sim 6000 \: km \: s^{-1}$) and of SN 2012A ($\sim 5200 \: km \: s^{-1}$). After $\sim 100$ days it drops to $\sim 3000 \: km \: s^{-1}$ for SN 2012aw, to $\sim 2500 \: km \: s^{-1}$ for SN 2012ec and to $\sim 2000 \: km \: s^{-1}$ for SN 2012A.  In terms of ejecta velocities, SN 2012ec is intermediate between SN 2012aw and SN 2012A.

\begin{figure}
  \centering
  \includegraphics[scale=0.4]{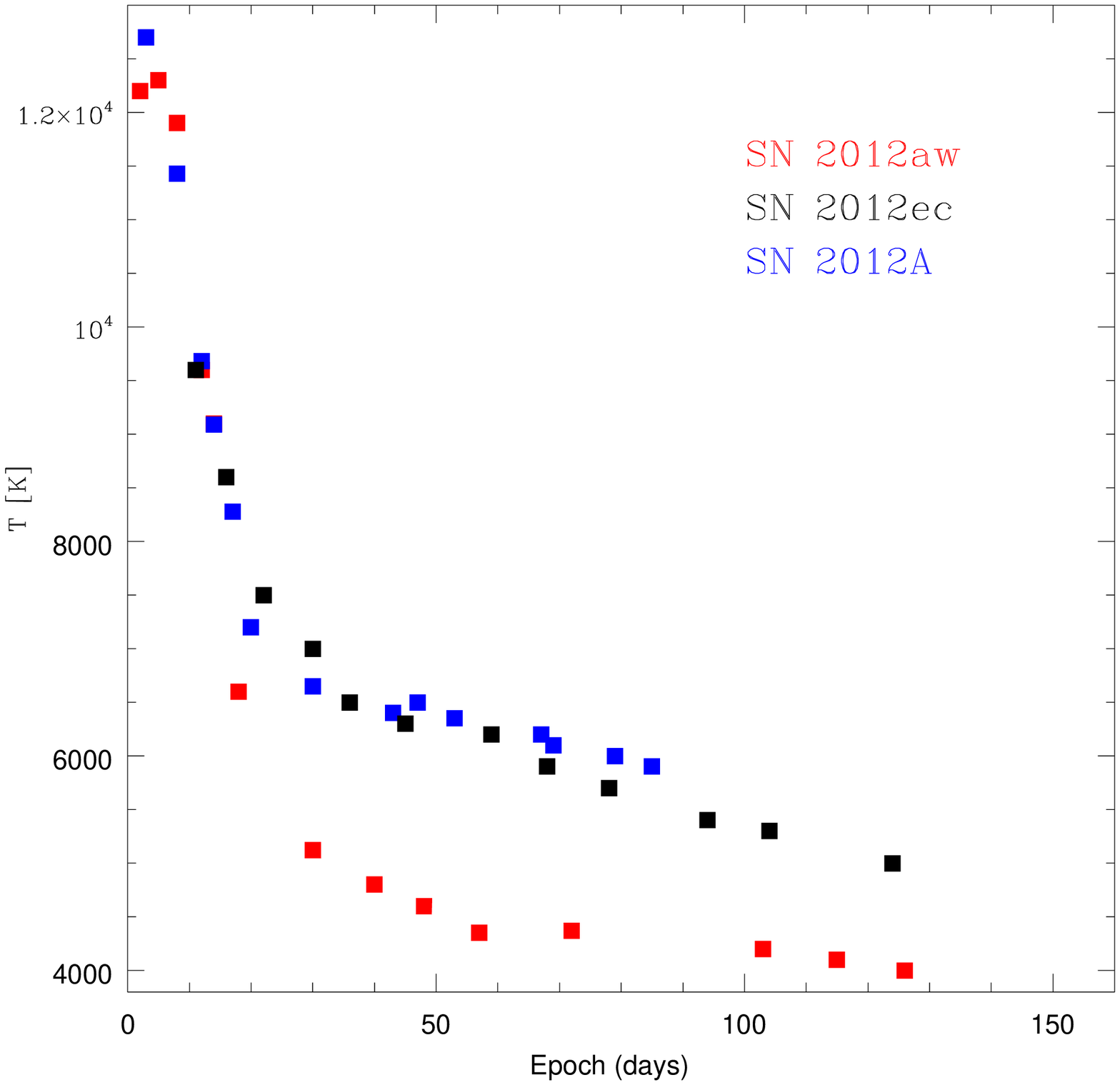}
  \caption{Comparison of the time evolution of the photospheric temperatures of SNe 2012ec, 2012A and 2012aw.}
  \label{temp}
\end{figure}

A comparison of the temperature estimated via blackbody fitting of the SED evolution for the 3 SNe is presented in Fig. \ref{temp}, from which it is clear that the temperature evolutions of SN 2012ec and SN 2012A are similar, and significantly hotter than SN 2012aw (from $\sim 20-30$ days post-explosion).

The ejected mass calculated for SN 2012ec is $12.6 \: \msun$, which is comparable to the value estimated for SN 2012A ($12.5 \: \msun$ \citealt{Tomasella2013}), but lower than value calculated for SN 2012aw ($ 20 \: \msun$, \citealt{DallOra2014}). Similarly the initial radius for SN 2012ec is comparable to SN 2012A ($\sim 260 \: R_{\odot}$), but smaller than for SN 2012aw ($\sim 400 \: R_{\odot}$).  Conversely, the estimated energy of SN 2012ec of $1.2 \: foe$ is higher than the value estimated for SN 2012A ($0.48 \: foe$) but similar to the energy of SN 2012aw ($1.5 \: foe$).

In summary, SN 2012ec is more luminous than SN 2012A, synthesised more $\mathrm{^{56}Ni}$ and has higher expansion velocities. The ejecta masses of the two SNe are comparable, but the pre-SN radius and the masses of the progenitors are slightly different.  This indicates that the progenitor of SN 2012ec progenitor was likely to be more massive, but more compact the progenitor of SN 2012A.  SN 2012aw has a larger initial radius, a more massive envelope and more energetic explosion that produced more $^{56}Ni$ and higher ejecta velocities than SN 2012ec. It is interesting to compare these estimates with the analysis of \citet{Poznanski2013}, who suggests a simple scaling relation between the energy deposited in the exploding core and the mass of the progenitor that, in turn, reflects on a linear correlation between mass and ejecta velocity. In particular, the positions of the ejected masses from the hydrodynamical code of SN 2012A and SN 2012aw in the Figure 1 of \citet{Poznanski2013}, are consistent with a steeper law $M \propto v^{1.5}$, while the ejected mass for SN 2012ec is much lower than expected from both the $M \propto v$ and $M \propto v^{1.5}$ relations. Since the hydrodynamical code estimates the ejecta masses, and not the progenitor masses, for SN 2012ec the discrepancy could be explained with a very efficient mass-loss mechanism. Unfortunately, the same argument cannot be invoked for SN 2012A and SN 2012aw. We also note that the \citet{Poznanski2013} analysis was based on progenitor masses estimated from stellar evolution models, which are based on a different input physics than the hydrodynamical codes.

The main characteristics of the comparisons between the three SNe are summarised in Table \ref{comp_SNe}.

\begin{table}
\caption{Comparison of the main parameters of SNe 2012ec, 2012aw and 2012A.}\label{comp_SNe}
\begin{footnotesize}
\begin{tabular}{lcccc}
\hline         
 &   SN  2012aw    & SN 2012ec &  SN 2012A \\
\hline 
$\mu$ (mag) & 29.96 & 31.19 & 29.96 \\
E(B-V) (mag) & 0.086 & 0.124 & 0.037 \\
$MJD_{expl}$ (d) & 56002 & 56151 & 55933 \\
$MJD_{disc}$ (d) & 56003 & 56143 & 55934 \\
$v_{Fe II}$ ($km \: s^{-1}$)$^{a}$ & $\sim 4200$ & $\sim 3700$ & $\sim 2800$ \\
$M_{R}$ (mag) & -17.1 & -16.7 & -16.2 \\
$L (10^{42} erg \: s^{-1})^{b}$ & 1.1 & 0.9 & 0.5 \\
Plateau duration (d) & 100 & 90 & 80 \\
$^{56}Ni$ ($\msun$) & 0.056 & 0.040 & 0.011 \\
E (foe)$^{c}$ & $ 1.5$ & 1.2 & 0.48 \\
R ($10^{13}$ cm) & 3  & 1.6 & 1.8 \\
$M_{eject}$ ($\msun$) & 20 & 12.6 & 12.5 \\
$M_{prog}$ ($\msun$)$^{d}$ & 13-16 & 14-22 & 8-15 \\
\hline
\end{tabular}
\\[1.5ex]
$^{a}$ at $\sim 50$ days \\
$^{b}$ at the plateau \\
$^{c}$ 1 foe= $10^{51} \: erg$ \\
$^{d}$ Mass of the progenitor as estimated from the pre-explosion images \\
\end{footnotesize}
\end{table}

\section{Type II-P SNe as Standard Candles}

The extragalactic distance scale is intimately connected with Type Ia SNe, up to cosmological distances, and through Type Ia SNe the acceleration of the Universe was discovered \citep{Perlmutter1999,Riess1998,Schmidt1998}. At the present time, current facilities allow us to detect and study Type Ia SNe up to $z=1.7$ \citep{Rubin2013}, while the next generation of extremely large telescopes will allow us to study Type Ia SNe up to $z\sim4$ \citep{Hook2013}. At high $z$, however, the number of Type Ia SNe may significantly decrease, due to the long lifetimes of their progenitors. Alternatively, the ubiquitous Type II (core-collapse) SNe could be an appealing choice to probe further cosmological distances. While Type Ia SNe are the product of an old to intermediate stellar population, Type II SNe come essentially from a young stellar population, and thus constitute a homogeneous sample with respect to the age of the stellar population.  It should also be noted, however, that type II SNe are significantly fainter than Type Ia SNe and that they explode in younger and dustier regions, making their discovery and study more difficult.

Although the characteristics of the light curves of the Type II SNe (peak luminosity, decline rate, presence and duration of the plateau) span a broad range of values, their use as distance indicators was already recognized by \citet{Kirshner1974}, who applied the Baade-Wesselink analysis to SN 1969L and SN 1970G through the Expanding Photosphere Method (EPM), and by \citet{Mitchell2002}, who modified the EPM method by introducing spectral synthesis analysis (Spectral-fitting Expanding Atmosphere Method, SEAM). Subsequently, \citet{Dessart2005} further exploited the EPM method by applying non-LTE atmospheric models. Both EPM and SEAM have been succesfully applied to SNe at cosmological distances (e.g. \citealt{Baron2004}, \citealt{Schmidt1994}), but require well sampled light curves and high quality spectra.

More specifically, for type II-P SNe, \citet{Hamuy2002} found a tight empirical correlation between the bolometric luminosity and the expansion velocity of the ejecta during the plateau phase. The luminosity and the expansion velocity (as measured from the Fe~II (5169\AA) line) are estimated at approximately the ``half plateau'' phase, conventionally set at $50$ days.  This method, dubbed the ``Standardized Candle Method'' (SCM), was subsequently investigated by \citet{Nugent2006}, \citet{Poznanski2009}, \citet{DAndrea2010} and \citet{Olivares2010}, with the advantage that it requires less input data than both EPM and SEAM. The empirical correlation at the base of the SCM was theoretically reproduced by \citet{Kasen2009}, who pointed out that the correlation relies on the simple behaviour of the expanding hydrogen envelope. They also warned, however, that the SCM may be sensitive to the progenitor metallicity and mass, that in turn could lead to systematic effects.

Almost all the quoted calibrations adopt  $50$ days post-explosion as a reference phase that roughly corresponds to the ``half-plateau''.  Other choices for the reference phase during the plateau phase can be set, but with the \textit{caveat} that the velocity measured from the Fe~II (5169) line is moderately decreasing over the duration of the plateau and that the method requires knowledge of the epoch of the explosion. Only \citet{Olivares2010} adopted a ``custom'' reference phase for each SN, due to the fact that the length of the plateau varies from SN to SN. For this reason, they suggested adopting a reference epoch $30$ days prior to the epoch at which the light curve has declined to a brightness midway between the plateau brightness and the brightness at which it joins the radioactive tail.

In this paper we take advantage of the homogeneous analysis of the three type II-P SNe (SNe 2012ec, 2012aw and 2012A) to perform a detailed comparison of the available calibrations of SCM and assess the robustness of the method. More specifically, for the comparison we adopt the $I$-band calibrations of SCM, namely: eq. $2$ of \citet{Hamuy2002}; eq.$1$ of \citet{Nugent2006}; eq. $2$ of \citet{Poznanski2009}; eq. 2 of \citet{DAndrea2010}\footnote{In passing, we note that the \citet{Poznanski2010} recalibration of this work led to a Hubble diagram with a scatter of only $11 \%$}; and eq. $16$ of \citet{Olivares2010}. Our estimated distances to the three SNe are compared with a homogeneous set of distances, based on primary (Cepheids, Tip of the Red Giant Branch, or TRGB) and secondary distance indicators (Tully-Fisher, Surface Brightness Fluctuations or SBF), available in the Extragalactic Distance Database \citep{Tully2009}. In Table \ref{distances} we report, for each SN, the distance estimated with the above calibrations. Moreover, we show the difference between the SCM distance and the estimates from the primary (when available) and secondary distance indicators. Finally, for each calibration, we report the mean difference and dispersion of the SCM distances with the estimates based on the primary and secondary distance indicators.

\begin{figure}
  \centering
  \includegraphics[width=0.5\textwidth ]{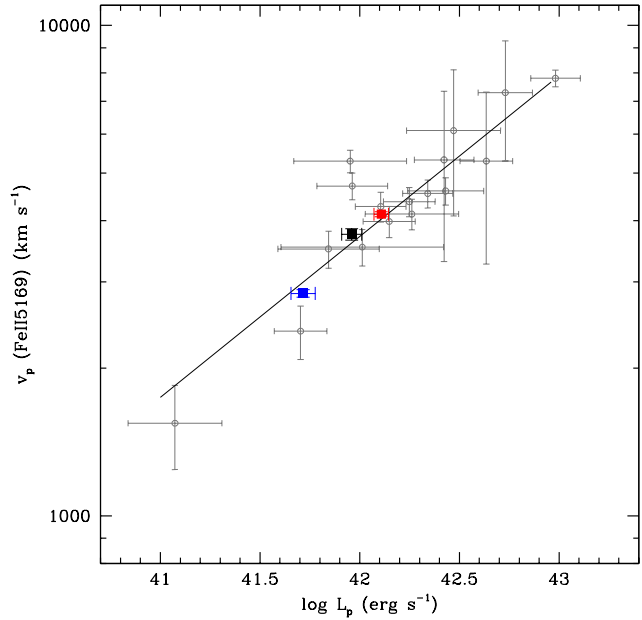}
  \caption{Our studied sample of type II-P SNe: SN 2012ec (black), SN 2012aw (red) and SN 2012A (blue) in the original \citet{Hamuy2002} plane.}
  \label{distance}
\end{figure}

\begin{table*}
\caption{Comparison of the SCM distances and the estimates from the primary and secondary distance indicators.\label{distances}}
\begin{tabular}{llcccrrr}
\hline
Calibration  &SN & SCM & Primary & Secondary & SCM $-$ Primary& SCM - Secondary & Mean residual\\ 
 & &(mag)& (mag)& (mag)& (mag)&(mag)& (mag) \\
\hline
             & SN 2012ec & $31.22 \pm 0.3$ &         & $31.19$ &        & $0.03$  &               \\
HP2002       & SN 2012aw & $29.96 \pm 0.3$ & $29.96$ & $30.00$ & $0.00$ & $-0.04$ & $0.01 \pm 0.04$\\
             & SN 2012A  & $30.05 \pm 0.3$ &         & $30.00$ &        & $0.05$  &                \\
\hline
             & SN 2012ec & $31.29 \pm 0.3$ &         & $31.19$ &        & $ 0.10$  &               \\
Nugent06     & SN 2012aw & $30.03 \pm 0.3$ & $29.96$ & $30.00$ & $0.07$ & $ 0.03$ & $-0.03 \pm 0.14$\\
             & SN 2012A  & $29.77 \pm 0.3$ &         & $30.00$ &        & $-0.23$  &               \\
\hline
             & SN 2012ec & $31.15 \pm 0.2$ &         & $31.19$ &         & $-0.04$  &               \\
Poznanski09  & SN 2012aw & $29.70 \pm 0.2$ & $29.96$ & $30.00$ & $-0.26$ & $-0.30$ & $-0.1 \pm 0.14$\\
             & SN 2012A  & $30.04 \pm 0.2$ &         & $30.00$ &         & $0.04$  &                \\
\hline
             & SN 2012ec & $31.11 \pm 0.2$ &         & $31.19$ &         & $-0.08$  &                 \\
Olivares10   & SN 2012aw & $29.58 \pm 0.2$ & $29.96$ & $30.00$ & $-0.38$ & $-0.42$  & $-0.01 \pm 0.37$\\
             & SN 2012A  & $30.47 \pm 0.2$ &         & $30.00$ &         & $0.47$   &                 \\
\hline
             & SN 2012ec & $31.33 \pm 0.2$ &         & $31.19$ &         & $0.14 $  &                 \\
D'Andrea10   & SN 2012aw & $29.86 \pm 0.2$ & $29.96$ & $30.00$ & $-0.10$ & $-0.14$  & $0.09 \pm 0.17$\\
             & SN 2012A  & $30.27 \pm 0.2$ &         & $30.00$ &         & $0.27$   &                 \\
\hline
\end{tabular}
\\[1.5ex]
Quoted errors for the SCM distances are the standard deviations of the individual calibrations. The value of the distance from the primary indicators of SN 2012aw is the average from the Cepheids \citep{Freedman2001} and the TRGB \citep{Rizzi2007} estimates. Finally, the ``mean residual'' column shows the average of the SCM $-$ Secondary values, where the error is the standard deviation.
\end{table*}

Table \ref{distances} may suggest that the \citet{Hamuy2002} calibration gives more homogenous results with respect to other calibrations. However, it must be noted that our test is based on only three SNe and that all the calibrations are consistent within the errors. We note that the \citet{Hamuy2002} calibration was derived assuming a value of $H_0 = 65$ km s$^{-1}$ Mpc$^{-1}$, significantly lower than the estimate of $H_0 = 73.8 \pm 2.4$ km s$^{-1}$ Mpc$^{-1}$ of \citet{Riess2011}, but in agreement with $H_0 = 63.7 \pm 2.3$ km s$^{-1}$ Mpc$^{-1}$ given by \citet{Tammann2013}. The large scatter in the \citet{Olivares2010} calibration could be due to the difficulty in estimating the reference phase, when a well sampled light curve covering the end of the plateau, is not available. All these calibrations rely on moderately distant SNe, embedded in the Hubble flow or for which SBF distances are available. However, these distances could still be affected by systematics not completely understood. For these reasons a new calibration of the SCM, based on nearby type II-P SNe for which primary (Cepheids and TRGB) and homogenous secondary indicators (TRGB) distances are available, would be of great interest. Moreover, for these SNe the metallicity effects suggested by \citet{Kasen2009} could also be investigated.  The average of the five individual estimates of the distances for SN 2012ec gives a distance modulus of $31.22 \pm 0.08$ mag, which we adopt as our final SCM-based distance. This value is in excellent agreement with the Tully-Fisher distance of $31.19 \pm 0.13$, adopted for our analysis.

\section{Conclusions}

We have presented the results of the Large Program ``Supernova Variety and Nuclesosynthesis Yelds'' and PESSTO photometric and spectroscopic monitoring campaign of SN 2012ec. This is one of the most intensively observed and well investigated Type II-P SNe to date. The optical and spectrocopic monitoring during the photospheric phase lasted for $\sim 161$ days and allowed us to determine the evolution of the pseudo-bolometric luminosity, the expansion velocity and the photospheric temperature and $^{56}Ni$ mass.
These parameters, analysed in conjunctions with hydrodynamical models, allowed us to estimate the explosion parameters such as the explosion energy, the envelope mass and the pre-SN radius.
Correcting the data for reddening ($E(B-V)=0.14 \pm ^{+0.15}_{-0.12}$ mag) and distance modulus ($\mu = 31.19 \pm 0.13 $) we estimated the luminosity to be $L= 0.9 \times 10^{42} \: erg \: s^{-1}$, at the plateau and evaluated the $^{56}Ni$ mass to be $0.040 \pm 0.015 \: \msun$.
The spectra of SN 2012ec were dominated by Balmer lines in the early epochs and after 20 days the iron-group elements started to appear and become more prominent with time. The NIR spectra were dominated by Paschen lines and, starting from 68 days, it is possible to identify He I, Ca I and $Br_{\gamma}$. A black body fit to the continuum gives temperatures of $11900 \pm 900$ K in the early epoches decreasing to $6200 \pm 500$ K at 50 days and $5000 \pm 500$ K in the last epochs.
From the spectroscopic dataset we estimate an initial velocity of $12200 \: km \: s^{-1}$ for the $H_{\alpha}$ line and $11000 \: km \: s^{-1}$ for $H_{\beta}$. The $H_{\alpha}$ velocity decreases to $5000 \; km \: s^{-1}$ by 50 days. At $\sim 25$ days the iron-group elements appear, for which we measure a velocity of $6000 \: km \: s^{-1}$ (for Fe II). The behaviour of SN 2012ec is similar to that seen in other II-P SNe, such as SN 1999em \citep{Elmhamdi2003} and SN 2004et \citep{Maguire2010}.\\

We estimate the physical parameters of SN 2012ec through the hydrodynamical modeling described in Sect. \ref{modelling}. The fit suggests an ejected mass of $M_{env}= 12.6 \: \msun$, a pre-SN radius of $R= 1.6 \times 10^{13} \: cm$, an explosion energy of $E=1.2 \: foe$ and an ejected $M(^{56}Ni)= 0.035 \: \msun$. The progenitor mass is in agreement with independent estimate of \citet{Maund2013} $M= 14-22 \: \msun$ obtained by analyzing pre-explosion images and of \citet{Jerkstrand14b}, submitted, $M= 13-15 \: \msun$ obtained from modeling of the spectra in the nebular phase.
Previously reported ejecta masses estimated from hydrodynamical modelling are generally too large compared to the initial mass estimated from direct detections of the progenitor on pre-explosion images \citep{Utrobin2008,Maguire2010}.  In order to investigate this discrepancy, we performed an homogeneous comparison between three  type II-P SNe, estimating the mass of the progenitor with two different approaches.
The methods and the codes used for the three objects in both cases are the same, to facilitate a reliable comparison.
We analyze the bright SN 2012aw \citep{DallOra2014}, the low-luminosity SN 2012A \citep{Tomasella2013} and SN 2012ec.
Several observational and derived parameters have been compared for these three objects.
SN 2012aw ($M_{R}=-17.1$ mag, at plateau) is brighter then SN 2012ec ($M_{R}=-16.7$ mag), while SN 2012A is fainter ($M_{R}=-16.2$ mag). A comparison between the bolometric light curves shows that SN 2012ec has an intermediate luminosity between the high luminosity SN 2012aw and the fainter SN 2012A. The nickel mass synthetized by these SNe is $M(^{56}Ni)_{12aw}= 0.056 \pm 0.013 \: \msun$, $M(^{56}Ni)_{12ec}= 0.040 \pm 0.015 \: \msun$ and $M(^{56}Ni)_{12A}= 0.011 \pm 0.004 \: \msun$.
A spectroscopic comparison shows a similar time evolution at all epochs. The velocities of $H_{\alpha}$, $H_{\beta}$ and Fe II of SN 2012ec, place it in the middle of the higher velocities from SN 2012aw and the slowest SN 2012A at all times.
The temperatures estimated are comparable for the three objects within the first 20 days, rather SN 2012ec tend to be similar to SN 2012A and they both are hotter than SN 2012aw.
SN 2012aw has a more energetic explosion ($E=1.5$ foe) than SN 2012ec and SN 2012A ($E= 0.48$ foe), but SN 2012ec is also more energetic than SN 2012A.
We finally compared the results of the direct detection of the progenitors of these three SNe with the masses estimated from the hydrodynamical modelling.
The progenitor mass estimated for SN 2012aw from the pre-explosion images ($M=13-16 \; \msun$) and from the hydrodynamical modeling ($M_{eject}= 20 \: \msun$) show that the two methods are not in good agreement and that SN 2012aw has a more massive progenitor then SN 2012ec, the last one having comparable ejecta mass with SN 2012A ($M= 8-15 \: \msun$, $M_{eject}= 12.5 \: \msun$).
The estimated initial radius of SN 2012aw ($R= 3 \times 10^{13}$ cm) indicate a larger progenitor then for SN 2012ec and SN 2012A ($R= 1.8 \times 10^{13}$ cm).
The estimates of the initial radius from the hydrodynamical modelling  for the three objects is lower than those from the pre-explosion images and seem to be too low for a RSG progenitor.
This homogeneous analysis finds a substantial match, within the errors, of the mass of the progenitor obtained with the two methods, mitigating the discrepancy which was pointed out in previous works \citep{Maguire2010}.
SN 2012ec, SN 2012aw and SN 2012A also follow the relation obtained by \citealt{Hamuy2002}.
This fact, coupled with their high luminosity at UV wavelengths, make Type II-P SNe interesting probes observable with the next generation of telescopes up to high \textit{z}.

\section{Acknowledgements}

We warmly thank our referee, Dovi Poznanski, for his helpful comments, which significantly improved the content and the readability of our manuscript.

We thank E. Cappellaro for the useful discussions.

C.B. thanks the IRAP PhD program for the financial support.
The research of JRM is supported through a Royal Society Research Fellowship.
A.G.-Y. is supported by an EU/FP7-ERC grant no [307260], "The Quantum Universe" I-Core program by the Israeli Committee for planning and budgeting and the ISF, GIF, Minerva, ISF and Weizmann-UK grants, and the Kimmel award.
G.P. acknowledges partial support by proyecto interno UNAB DI-303-13/R.
GP and MH  acknowledge support provided by the Millennium Institute of Astrophysics (MAS) through grant IC120009 of the Programa Iniciativa Cientifica Milenio del Ministerio de Economia, Fomento y Turismo de Chile".
M.D.V., M.L.P., S.B., A.P., L.T. and M.T. are partially supported by the PRIN-INAF 2011 with the project ``Transient Universe: from ESO Large to PESSTO".
This work was partly supported by the European Union FP7 programme through ERC grant number 320360.

This work is based (in part) on observations collected at the European Organisation for Astronomical Research in the Southern Hemisphere, Chile as part of PESSTO, (the Public ESO Spectroscopic Survey for Transient Objects Survey) ESO program 188.D-3003, 191.D-0935.
The research leading to these results has received funding from the European Research Council under the European Union's Seventh Framework Programme (FP7/2007-2013)/ERC Grant agreement n$^{\rm o}$ [291222]  (PI : S. J. Smartt) and STFC grants ST/I001123/1 and ST/L000709/1.

The early SN 2012ec data have been collected via the ESO-NTT Large Program “Supernova Variety and Nuclesosynthesis Yelds” (184.D-1140), an European supernova collaboration led by Stefano Benetti (http://sngroup.oapd.inaf.it/esolarge.html).
This paper is partially based on observations collected at Copernico telescope (Asiago, Italy) of the INAF - Osservatorio Astronomico di Padova; at the Galileo 1.22m Telescope operated by Department of Physics and Astronomy of the University of Padova at Asiago; at the 2.56m Nordic Optical Telescope operated by The Nordic Optical Telescope Scientific Association (NOTSA); at the 4.3m William Herschel Telescope operated by the Isaac Newton Group of Telescopes; on observations obtained through the CNTAC proposal CN2012B-73 and on observations made with the Liverpool Telescope (programme OL12B) operated on the island of La Palma by Liverpool John Moores University in the Spanish Observatorio del Roque de los Muchachos of the Instituto de Astrofisica de Canarias with financial support from the UK Science and Technology Facilities Council.


\begin{thebibliography}{}

\bibitem[\protect\citeauthoryear{{Arcavi}, {Gal-Yam}, {Cenko} et al.}{{Arcavi et al.}}{2012}]{Arcavi2012} {Arcavi} I., {Gal-Yam} A., {Cenko} S.B. et al., 2012, \apj, 756L, 30A

\bibitem[\protect\citeauthoryear{{Arnett} $\&$ {Fu}}{{Arnett $\&$ Fu}}{1989}]{Arnett1989}
{Arnett} W.D. $\&$ {Fu} A. 1989, \apj, 340, 396
 
\bibitem[\protect\citeauthoryear{{Barbon}, {Ciatti} $\&$ {Rosino}}{Barbon et al.}{1979}]{Barbon1979}
{Barbon} R., {Ciatti} F., \& {Rosino} L., 1979, \aap, 72, 287 

\bibitem[\protect\citeauthoryear{{Baron}, {Nugent}, {Branch} et al.}{{Baron et al.}}{2004}]{Baron2004} {Baron} E., {Nugent} P.E., {Branch} D. et al., 2004, \apj, 616, L91

\bibitem[\protect\citeauthoryear{{Bayless}, {Pritchard}, {Roming} et al.}{{Bayless et al.}}{2013}]{Bayless2013}
{Bayless} A.J., {Pritchard} T.A., {Roming} P.J., et al., 2013, \apj, 764L, 13B

\bibitem[\protect\citeauthoryear{{Bersten}, {Benvenuto}, {Nomoto} et al.}{{Bersten et al.}}{2012}]{Bersten2012}
{Bersten} M.C., {Benvenuto} O., {Nomoto} K. et al., 2012, \apj, 757, 31B

\bibitem[\protect\citeauthoryear{{Blinnikov}, {Lundqvist}, {Bartunov} et al.}{{Blinnikov et al.}}{2000}]{Blinnikov2000}
{Blinnikov} S., {Lundqvist} P., {Bartunov} O., et al. 2000, \apj, 532, 1132B

\bibitem[\protect\citeauthoryear{{Botticella}, {Trundle}, {Pastorello} et al.}{{Botticella et al.}}{2012}]{Botticella2010}
{Botticella} M.T., {Trundle} C., {Pastorello} A. et al., 2010, ApJ, 717L, 52B

\bibitem[\protect\citeauthoryear{{Botticella}, {Kumar}, {Sutaria} et al.}{{Bose et al.}}{2013}]{Bose2013}
{Bose} S., {Kumar} B., {Sutaria} F. et al., 2013, \mnras, 433, 1871B

\bibitem[\protect\citeauthoryear{{Bose}, {Smartt}, {Kennicutt} et al.}{{Botticella et al.}}{2012}]{Botticella2012}
{Botticella} M.T., {Smartt} S.J., {Kennicutt} R.C. et al., 2012, A$\&$A, 537A, 132B

\bibitem[\protect\citeauthoryear{{Bouchet}, {Danziger} $\&$ {Lucy}}{{Bouchet et al.}}{1991}]{Bouchet1991}
{Bouchet} P., {Danziger} I.J. $\&$ {Lucy} L.B. 1991, \aj, 102, 1135

\bibitem[\protect\citeauthoryear{{Cappellaro}, {Evans} $\&$ {Turatto}}{{Cappellaro et al.}}{1999}]{Cappellaro1999}
{Cappellaro} E., {Evans} R. $\&$ {Turatto} M. 1999, \aap, 351, 459

\bibitem[\protect\citeauthoryear{{Cardelli}, {Clayton} $\&$ {Mathis}}{{Cardelli et al.}}{1989}]{Cardelli1989}
{Cardelli} J.A., {Clayton} G.C. $\&$ {Mathis} J.S. 1989, \apj, 345, 245

\bibitem[\protect\citeauthoryear{{Carpenter}}{{Carpenter}}{2001}]{Carpenter2001}
{Carpenter} J.M., 2001, \aj, 121, 2851

\bibitem[\protect\citeauthoryear{{Clocchiatti}, {Benetti}, {Wheeler} et al.}{{Clocchiatti et al.}}{1996}]{Clocchiatti1996}
{Clocchiatti} A., {Benetti} S., {Wheeler} J.C., et al. 1996, AJ, 111, 1286C

\bibitem[\protect\citeauthoryear{{Coppola}, {Dall'Ora}, {Ripepi} et al.}{{Coppola et al.}}{2011}]{Coppola2011}
{Coppola} G., {Dall'Ora} M., {Ripepi} V., et al. 2011, \mnras, 416, 1056C

\bibitem[\protect\citeauthoryear{{Childress} {Scalzo} {Yuan} {Schmidt}}{{Childress et al.}}{2012}]{Childress2012}
{Childress} M., {Scalzo} R., {Yuan} F., {Schimdt} B., 2012, Central Bureau Electronic Telegrams, 3201, 2

\bibitem[\protect\citeauthoryear{{Chugai} $\&$ {Utrobin}}{{Chugai $\&$ Utrobin}}{2000}]{Chugai2000}
{Chugai} N.N., $\&$ {Utrobin} V.P., 2000, A$\&$A, 354, 557C

\bibitem[\protect\citeauthoryear{{D'Andrea}, {Sako}, {Dilday} et al.}{{D'Andrea et al.}}{2010}]{DAndrea2010}
{D'Andrea} C.B., {Sako} M., {Dilday} B., et al. 2010, \apj, 708, 661D

\bibitem[\protect\citeauthoryear{{Dalhen}, {Strolger}, {Riess} et al.}{{Dalhen et al.}}{2012}]{Dalhen2012}
{Dalhen} T., {Strolger} L.G., {Riess} A.G., et al. 2012, \apj, 757, 70D

\bibitem[\protect\citeauthoryear{{Dall'Ora}, {Botticella}, {Pumo} et al.}{{Dall'Ora et al.}}{2014}]{DallOra2014}
{Dall'Ora} M., {Botticella} M.T., {Pumo} L., et al. 2014, \apj, 787, 139

\bibitem[\protect\citeauthoryear{{Dessart} $\&$ {Hillier}}{{Dessart $\&$ Hillier}}{2005}]{Dessart2005}
{Dessart} L., $\&$ {Hillier} D.J., 2005, A$\&$A, 439, 671

\bibitem[\protect\citeauthoryear{{Eastman}, {Schmidt} $\&$ {Kirshner}}{{Eastman, Schmidt $\&$ Kirshner}}{1996}]{Eastman1996}
{Eastman} R.G., {Schmidt} B.P. $\&$ {Kirshner} R., 1996, ApJ, 466, 911E

\bibitem[\protect\citeauthoryear{{Eldridge} $\&$ {Tout}}{{Eldridge $\&$ Tout}}{2004}]{Eldridge2004}
{Eldridge} J.J. $\&$ {Tout} C.A., 2004, \mnras, 353, 87E

\bibitem[\protect\citeauthoryear{{Elmhamdi}, {Danziger}, {Chugai}}{{Elmhamdi et al.}}{2003}]{Elmhamdi2003}
{Elmhamdi} A., {Danziger} I.J., {Chugai} N. et al., 2003, \mnras, 338, 939E

\bibitem[\protect\citeauthoryear{{Elmhamdi}, {Chugai} $\&$ {Danziger}}{{Elmhamdi, Chugai $\&$ Danzinger}}{2003}]{Elmhamdi2003b}
{Elmhamdi} A.,{Chugai} N., $\&$ {Danziger} I.J., 2003, A$\&$A, 404, 1077E

\bibitem[\protect\citeauthoryear{{Faran}, {Poznanski}, {Filippenko} et al.}{{Faran et al.}}{2014}]{Faran2014}
{Faran} T., {Poznanski} D., {Filippenko} A.V., et al., 2014, \mnras, 442, 844F

\bibitem[\protect\citeauthoryear{{Filippenko}}{{Filippenko}}{1997}]{Filippenko1997}
{Filippenko} A.V., 1997, ARA$\&$A, 35, 309

\bibitem[\protect\citeauthoryear{{Fraser}, {Maund}, {Smartt} et al.}{{Fraser et al.}}{2012}]{Fraser2012}
{Fraser} M., {Maund} J.R., {Smartt} S.J., et al., 2012, \apj, 759L, 13F

\bibitem[\protect\citeauthoryear{{Freedman}, {Madore}, {Gibson} et al.}{{Freedman et al.}}{2001}]{Freedman2001}
{Freedman} W.L., {Madore} B.F., {Gibson} B.K., et al., 2001, \apj, 553, 47F

\bibitem[\protect\citeauthoryear{{Gilmozzi}, {Cassatella}, {Clavel}}{{Gilmozzi et al.}}{1987}]{Gilmozzi1987}
{Gilmozzi} R., {Cassatella} A., {Clavel} J., et al., 1987, Nature, 328, 318G

\bibitem[\protect\citeauthoryear{{Grassberg}, {Imshennik}, {Nadyozhin}}{{Grassberg et al.}}{1971}]{Grassberg1971}
{Grassberg} E.K., {Imshennik} V.S., {Nadyozhin} D.K., 1971, Ap$\&$SS, 10, 28G

\bibitem[\protect\citeauthoryear{{Gustafsson}, {Edvardsson}, {Eriksson}, et al.}{{Gustafsson et al.}}{2008}]{Gustafsson2008}
{Gustafsson} B., {Edvardsson} B., {Eriksson} K., 2008, A$\&$A, 486, 951G

\bibitem[\protect\citeauthoryear{{Hamuy} et al.}{{Hamuy et al.}}{2001}]{Hamuy2001}
{Hamuy} M. et al., 2001, \apj, 566, L63

\bibitem[\protect\citeauthoryear{{Hamuy} $\&$ {Pinto}}{{Hamuy $\&$ Pinto}}{2002}]{Hamuy2002}
{Hamuy} M. $\&$ {Pinto} P., 2002, \apj, 558, 615

\bibitem[\protect\citeauthoryear{{Harutyunyan}, {Pfahler}, {Pastorello}, et al.}{{Harutyunyan et al.}}{2008}]{Harutyunyan2008}
{Harutyunyan} A., {Pfahler} P., {Pastorello} A., 2008, A$\&$A, 488, 383H

\bibitem[\protect\citeauthoryear{{Heger}, {Fryer}, {Woosley}, {Langer}, $\&$ {Hartmann}}{{Heger et al.}}{2003}]{Heger2003}
{Heger} A., {Fryer} C.L., {Woosley} S.E., {Langer} N., $\&$ {Hartmann} D.H., 2003, \apj, 591, 288

\bibitem[\protect\citeauthoryear{{Hook}}{{Hook}}{2013}]{Hook2013}
{Hook} I.M., 2013, Royal Society of London Philosophical Transactions Series A, 371, 20282 

\bibitem[\protect\citeauthoryear{{Hopkins}, {Somerville}, {Hernquist}, et al.}{{Hopkins et al.}}{2006}]{Hopkins2006}
{Hopkins} P.F., {Somerville} R.S., {Hernquist} L., et al., 2006, \apj, 652, 864H

\bibitem[\protect\citeauthoryear{{Iben} $\&$ {Renzini}}{{Iben $\&$ Renzini}}{1983}]{Iben1983}
{Iben} I. $\&$ {Renzini} A., 1983, ARA$\&$A, 21, 271

\bibitem[\protect\citeauthoryear{{Inserra}, {Turatto}, {Pastorello} et al.}{{Inserra et al.}}{2011}]{Inserra2011}
{Inserra} C., {Turatto} M., {Pastorello} A. et al, 2011, \mnras, 417, 261I

\bibitem[\protect\citeauthoryear{{Inserra}, {Turatto}, {Pastorello} et al.}{{Inserra et al.}}{2012}]{Inserra2012}
{Inserra} C., {Turatto} M., {Pastorello} A. et al, 2012, \mnras, 422, 1122

\bibitem[\protect\citeauthoryear{{Jerkstrand}, {Fransson}, {Maguire} et al.}{{Jerkstrand et al.}}{2012}]{Jerkstrand2012}
{Jerkstrand} A., {Fransson} C., {Maguire} K. et al, 2012, A\&A, 546A, 28J

\bibitem[\protect\citeauthoryear{{Jerkstrand}, {Smartt}, {Fraser} et al.}{{Jerkstrand et al.}}{2014}]{Jerkstrand2014}
{Jerkstrand} A., {Smartt} S.J., {Fraser} M. et al, 2014, \mnras, 439, 3694J

\bibitem[Jerkstrand et al.(2014)]{Jerkstrand14b} Jerkstrand, A., Smartt, S.~J., Sollerman, J., et al.\ 2014, arXiv:1410.8394 

\bibitem[\protect\citeauthoryear{{Kasen} $\&$ {Woosley}}{{Kasen $\&$ Woosley}}{2009}]{Kasen2009}
{Kasen} D. $\&$ {Woosley} S.E., 2009, \apj, 703, 2205

\bibitem[\protect\citeauthoryear{{King}, {Modjaz}, {Shefler} et al.}{{King et al.}}{1998}]{King1998}
{King} J.Y., {Modjaz} M., {Shefler} T. et al, 1998, IAUC, 6992, 1

\bibitem[\protect\citeauthoryear{{Kirshner} $\&$ {Kwan}}{{Kirshner $\&$ {Kwan}}}{1974}]{Kirshner1974}
{Kirshner} R.P. $\&$ {Kwan} J., 1974, \apj, 193, 27K

\bibitem[\protect\citeauthoryear{{Kirshner}, {Sonneborn},{Crenshaw}}{{Kirshner et al.}}{1987}]{Kirshner1987}
{Kirshner} R.P., {Sonneborn} G., {Crenshaw} D.M., et al., 1987, \apj, 320, 602K

\bibitem[Kleiser et al.(2011)]{Kleiser2011} Kleiser, I.~K.~W., 
Poznanski, D., Kasen, D., et al.\ 2011, \mnras, 415, 372 

\bibitem[\protect\citeauthoryear{{Kochanek}, {Khan} $\&$ {Dai}}{{Kochanek et al.}}{2012}]{Kochanek2012}
{Kochanek} C.S., {Khan} R. $\&$ {Dai} X., 2012, \apj, 759, 20K

\bibitem[\protect\citeauthoryear{{Kowal}}{{Kowal}}{1968}]{Kowal1968}
{Kowal} C.T., 1968, \aj, 73, 1021K

\bibitem[\protect\citeauthoryear{{Landolt}}{{Landolt}}{1992}]{Landolt1992}
{Landolt} A.U., 1992, \aj, 104, 340L

\bibitem[\protect\citeauthoryear{{Leonard} et al.}{{Leonard et al.}}{2002}]{Leonard2002}
{Leonard} D.C. et al., 2002, PASP, 114, 35L

\bibitem[\protect\citeauthoryear{{Li}, {Cenko}, {Filippenko} et al.}{{Li et al.}}{2009}]{Li2009}
{Li} W., {Cenko} S.B. $\&$ {Filippenko} A.V., 2009, Central Bureau Electronic Telegrams, 1656, 1

\bibitem[\protect\citeauthoryear{{Li}, {Leaman}, {Chornock}, et al.}{{Li et al.}}{2011}]{Li2011}
{Li} W., {Leaman} J., {Chornork} R., et al., 2011, \mnras, 412, 1441

\bibitem[\protect\citeauthoryear{{Litvinova} $\&$ {Nadezhin}}{{Litvinova $\&$ {Nadezhin}}}{1983}]{Litvinova1983}
{Litvinova} I. Yu., {Nadezhin} D.K., 1983, Ap$\&$SS, 89, 89L

\bibitem[\protect\citeauthoryear{{Litvinova} $\&$ {Nadezhin}}{{Litvinova $\&$ {Nadezhin}}}{1985}]{Litvinova1985}
{Litvinova} I. Yu., {Nadezhin} D.K., 1985, Pis'ma Astron. Zh., 11, 351

\bibitem[\protect\citeauthoryear{{Maguire} et al.}{{Maguire et al.}}{2010}]{Maguire2010}
{Maguire} K., et al. 2010, \mnras, 404, 981

\bibitem[\protect\citeauthoryear{{Massey}, {Waterhouse} $\&$ {DeGioia-eastwood}}{{Massey et al.}}{2000}]{Massey2000}
{Massey} P., {Waterhouse} E. $\&$ {DeGioia-Eastwood} K., 2000, \aj, 119, 2214

\bibitem[\protect\citeauthoryear{{Massey}, {DeGioia-eastwood} $\&$ {Waterhouse}}{{Massey et al.}}{2001}]{Massey2001}
{Massey} P., {DeGioia-Eastwood} K. $\&$ {Waterhouse} E. 2001, \aj, 121, 1050

\bibitem[\protect\citeauthoryear{{Maund}, {Fraser}, {Smartt} et al.}{{Maund et al.}}{2013}]{Maund2013}
{Maund} J.R., {Fraser} M., {Smartt} S.J., et al. 2013, \mnras, 413L, 102M

\bibitem[\protect\citeauthoryear{{Mitchell}, {Baron}, {Branch} et al.}{{Mitchell et al.}}{2002}]{Mitchell2002}
{Mitchell} R.C., {Baron} E., {Branch} D., et al. 2002, ApJ, 574, 293M

\bibitem[\protect\citeauthoryear{{Moiseev}}{{Moiseev}}{2000}]{Moiseev2000}
{Moiseev} A.V., 2000, \aap, 363, 843M

\bibitem[\protect\citeauthoryear{{Monard}}{{Monard}}{2012}]{Monard2012}
{Monard} L.A.G., 2012, Central Bureau Electronic Telegrams, 3201, 1

\bibitem[\protect\citeauthoryear{{Mould}, {Huchra}, {Freedman} et al.}{{Mould et al.}}{2000}]{Mould2000}
{Mould} J.R., {Huchra} J.P., {Freedman} W.L., et al. 2000, \apj, 529, 786M

\bibitem[\protect\citeauthoryear{{Nakano}, {Aoki}, {Kushida} et al.}{{Nakano et al.}}{1996}]{Nakano1996}
{Nakano} S., {Aoki} M., {Kushida} Y. et al, 1996, IAUC, 6442, 1

\bibitem[\protect\citeauthoryear{{Nakano}}{{Nakano}}{2006}]{Nakano2006}
{Nakano} S., 2006, CBET, 470, 1N

\bibitem[\protect\citeauthoryear{{Nugent}, {Sullivan}, {Ellis} et al.}{{Nugent et al.}}{2006}]{Nugent2006}
{Nugent} P., {Sullivan} M., {Ellis} R., et al. 2006, \apj, 645, 841 

\bibitem[\protect\citeauthoryear{{Olivares}, {Hamuy}, {Pignata} et al.}{{Olivares et al.}}{2010}]{Olivares2010}
{Olivares} E.F., {Hamuy} M., {Pignata} G., et al. 2010, \apj, 715, 833

\bibitem[\protect\citeauthoryear{{Pastorello}, {Zampieri}, {Turatto} et al.}{{Pastorello et al.}}{2004}]{Pastorello2004}
{Pastorello} A., {Zampieri} L., {Turatto} M., et al. 2004, \mnras, 347, 74P

\bibitem[\protect\citeauthoryear{{Pastorello}, {Valenti}, {Zampieri} et al.}{{Pastorello et al.}}{2009}]{Pastorello2009}
{Pastorello} A., {Valenti} S., {Zampieri} L., et al. 2009, \mnras, 394, 2266P

\bibitem[\protect\citeauthoryear{{Pastorello}, {Crockett}, {Martin} et al.}{{Pastorello et al.}}{2009}]{Pastorello2009b}
{Pastorello} A., {Crockett} R.M., {Martin} R., et al. 2009, A$\&$A, 500, 1013P

\bibitem[\protect\citeauthoryear{{Pastorello}, {Pumo}, {Navasardyan} et al.}{{Pastorello et al.}}{2012}]{Pastorello2012}
{Pastorello} A., {Pumo} M.L., {Navasardyan} H., et al. 2012, A$\&$A, 537A, 141P

\bibitem[\protect\citeauthoryear{{Perlmutter}, {Aldering}, {Goldhaber}, et al.}{{Perlmutter et al.}}{1999}]{Perlmutter1999}
{Perlmutter} S.,  {Aldering} G., {Goldhaber} G., et al. 1999, \apj, 517, 565P

\bibitem[Poznanski(2013)]{Poznanski2013} Poznanski, D.\ 2013, \mnras, 436, 3224 

\bibitem[\protect\citeauthoryear{{Poznanski}, {Butler}, {Filippenko} et al.}{{Poznanski et al.}}{2009}]{Poznanski2009}
{Poznanski} D., {Butler} N., {Filippenko} A.V., et al. 2009, \apj, 694, 1067

\bibitem[Poznanski et al.(2010)]{Poznanski2010} Poznanski, D., Nugent, P.~E., \& Filippenko, A.~V.\ 2010, \apj, 721, 956 

\bibitem[\protect\citeauthoryear{{Poznanski}, {Prochaska} $\&$ {Bloom}}{{Poznanski et al.}}{2012}]{Poznanski2012}
{Poznanski} D., {Prochaska} J.X., $\&$ {Bloom} J.S. 2012, \mnras, 426, 1465P

\bibitem[\protect\citeauthoryear{{Pumo}, {Zampieri} $\&$ {Turatto}}{{Pumo et al.}}{2010}]{Pumo2010}
{Pumo} M.L., {Zampieri} L. $\&$ {Turatto} M., 2010, MSAIS, 14, 123

\bibitem[\protect\citeauthoryear{{Pumo} $\&$ {Zampieri}}{{Pumo $\&$ Zampieri}}{2011}]{Pumo2011}
{Pumo} M.L. $\&$ {Zampieri} L., 2011, \apj, 741, 41

\bibitem[\protect\citeauthoryear{{Quimby}, {Wheeler}, et al.}{{Quimby et al.}}{2007}]{Quimby2007}
{Quimby} R.M., {Wheeler} J.G., et al. 2007, \apj, 666, 1093

\bibitem[\protect\citeauthoryear{{Ramya}, {Sahu} $\&$ {Prabhu}}{{Ramya et al.}}{2007}]{Ramya2007}
{Ramya} S., {Sahu} D.K. $\&$ {Prabhu} T.P., et al. 2007, \mnras, 381, 511 

\bibitem[\protect\citeauthoryear{{Riess}, {Filippenko}, {Challis}}{{Riess et al.}}{1998}]{Riess1998}
{Riess} A.g., {Filippenko} A.V., {Challis} P., et al. 1998, AJ, 116, 1009R

\bibitem[\protect\citeauthoryear{{Riess}, {Macri}, {Casertano}}{{Riess et al.}}{2011}]{Riess2011}
{Riess} A.G., {Macri} L., {Casertano} S., et al. 2011, \apj, 730, 119R

\bibitem[\protect\citeauthoryear{{Rizzi}, {Tully}, {Makarov} et al.}{{Rizzi et al.}}{2007}]{Rizzi2007}
{Rizzi} L., {Tully} R.B., {Makarov} D., et al. 2007, \apj, 661, 815R

\bibitem[\protect\citeauthoryear{{Roy}, {Sutaria}, {Bose} et al.}{{Roy et al.}}{2014}]{Roy2014}
{Roy} R., {Sutaria} F., {Bose} S., et al. 2007, IAUS, 296, 116R

\bibitem[\protect\citeauthoryear{{Rubin}, {Knop}, {Rykoff} et al.}{{Rubin et al.}}{2013}]{Rubin2013}
{Rubin} D., {Knop} R.A., {Rykoff} E., et al. 2013, ApJ, 763, 35R

\bibitem[\protect\citeauthoryear{{Schlafly} $\&$ {Finkbeiner}}{{Schlafly $\&$ Finkbeiner}}{2011}]{Schlafly2011}
{Schlafly} E.F. $\&$ {Finkbeiner} D.P., 2011, \apj, 737, 103

\bibitem[Schlegel et al.(1998)]{Schlegel1998} Schlegel, D.~J., Finkbeiner, D.~P., \& Davis, M.\ 1998, \apj, 500, 525 

\bibitem[\protect\citeauthoryear{{Schmidt}, {Kirshner}, {Eastman}}{{Schmidt et al.}}{1994}]{Schmidt1994}
{Schmidt} B.P., {Kirshner} R.P., {Eastman} R.G., 1994, \apj, 432, 42S

\bibitem[\protect\citeauthoryear{{Schmidt}, {Suntzeff}, {Phillips}}{{Schmidt et al.}}{1998}]{Schmidt1998}
{Schmidt} B.P., {Suntzeff} N.B., {Phillips} M.M., 1998, \apj, 507, 46S

\bibitem[\protect\citeauthoryear{{Skrutskie}, {Cutri}, {Stiening} et al.}{{Skrutskie et al.}}{2006}]{Skrutskie2006}
{Skrutskie} M.F., {Cutri} R.M., {Stiening} R., et al. 2006, \aj, 131, 1163

\bibitem[\protect\citeauthoryear{{Smartt}}{{Smartt}}{2009}]{Smartt2009}
{Smartt} S.J., 2009, ARA$\&$A, 47, 63

\bibitem[\protect\citeauthoryear{{Smartt}, {Eldridge}, {Crockett}, et al.}{{Smartt et al.}}{2009}]{Smarttetal2009}
{Smartt} S.J., {Eldridge} J.J., {Crockett} R.M., 2009, \mnras, 395, 1409S

\bibitem[Smartt et al.(2014)]{Smartt2014} Smartt, S.~J., Valenti, S., Fraser, M., et al.\ 2014, arXiv:1411.0299 

\bibitem[\protect\citeauthoryear{{Smith}, {Tucker}, {Kent}, et al.}{{Smith et al.}}{2002}]{Smith2002}
{Smith} J.A., {Tucker} D.L., {Kent} S., 2002, \aj, 123, 2121S

\bibitem[\protect\citeauthoryear{{Spiro}, {Pastorello}, {Pumo}, et al.}{{Spiro et al.}}{2014}]{Spiro2014}
{Spiro} S, {Pastorello} A., {Pumo} M.L., 2014, \mnras, 439, 2873S

\bibitem[\protect\citeauthoryear{{Stetson}}{{Stetson}}{1987}]{Stetson1987}
{Stetson} P.B., 1987, PASP, 99, 191S

\bibitem[\protect\citeauthoryear{{Suntzeff}, {Hamuy}, {Martin}, et al.}{{Suntzeff et al.}}{1988}]{Suntzeff1988}
{Suntzeff} N.B., {Hamuy} M., {Martin} G. et al., 1988, \aj, 96, 1864S

\bibitem[\protect\citeauthoryear{{Tammann} $\&$ {Reindl}}{{Tammann} $\&$ {Reindl}}{2013}]{Tammann2013}
{Tammann} G.A. $\&$ {Reindl} B., 2013, A$\&$A, 549A, 136T

\bibitem[\protect\citeauthoryear{{Tomasella}, {Cappellaro}, {Fraser} et al.}{{Tomasella et al.}}{2013}]{Tomasella2013}
{Tomasella} L., {Cappellaro} E., {Fraser} M., et al. 2013, \mnras, 434, 1636T

\bibitem[\protect\citeauthoryear{{Tully}, {Rizzi}, {Shaya} et al.}{{Tully et al.}}{2009}]{Tully2009}
{Tully} R., {Rizzi} L., {Shaya} E.J., et al. 2009, \aj, 138, 323T

\bibitem[\protect\citeauthoryear{{Turatto}, {Benetti}, {Cappellaro}}{{Turatto et al.}}{2003}]{Turatto2003}
{Turatto} M., {Benetti} S., {Cappellaro} E. 2003, in From Twilight to Highlight: The Physics of Supernovae, ed. W. Hillebrandt $\&$ B. Leibundgunt, 200

\bibitem[\protect\citeauthoryear{{Utrobin}}{{{Utrobin}}}{1993}]{Utrobin1993}
{Utrobin} V.P., 1993, A$\&$A, 281L, 89U

\bibitem[\protect\citeauthoryear{{Utrobin}}{{{Utrobin}}}{2007}]{Utrobin2007a}
{Utrobin} V.P., 2007, A$\&$A, 461, 233U

\bibitem[\protect\citeauthoryear{{Utrobin}, {Chugai}, {Pastorello}}{{{Utrobin} et al.}}{2007}]{Utrobin2007}
{Utrobin} V.P., {Chugai} N.N., {Pastorello} A. 2007, A$\&$A, 475, 973U

\bibitem[\protect\citeauthoryear{{Utrobin} $\&$ {Chugai}}{{{Utrobin} $\&$ {Chugai}}}{2008}]{Utrobin2008}
{Utrobin} V.P. $\&$ {Chugai} N.N., 2008, A$\&$A, 491, 507U

\bibitem[\protect\citeauthoryear{{Utrobin} $\&$ {Chugai}}{{{Utrobin} $\&$ {Chugai}}}{2009}]{Utrobin2009}
{Utrobin} V.P. $\&$ {Chugai} N.N., 2009, A$\&$A, 506, 829U

\bibitem[\protect\citeauthoryear{{Valenti}, {Fraser}, {Benetti} et al.}{{Valenti et al.}}{2011}]{Valenti2011}
{Valenti} S., {Fraser} M., {Benetti} S., et al. 2011, \mnras, 416, 3138

\bibitem[\protect\citeauthoryear{{Van Dyk}, {Cenko}, {Poznanski} et al.}{{Van Dyk et al.}}{2012}]{Van2012}
{Van Dyk} S.D., {Cenko} S.B., {Poznanski} D., et al. 2012, \apj, 756, 131V

\bibitem[\protect\citeauthoryear{{Walmswell} $\&$ {Eldridge}}{{Walmswell $\&$ Eldridge}}{2012}]{Walmswell2012}
{Walmswell} J.J. $\&$ {Eldridge} J.J., 2012, \mnras, 419, 2054

\bibitem[\protect\citeauthoryear{{Weaver}, $\&$ {Woosley}}{{Weaver $\&$ {Woosley}}}{1980}]{Weaver1980}
{Weaver} T.A., $\&$ {Woosley} S.E., 1980, AIPC, 63, 15W

\bibitem[\protect\citeauthoryear{{Woosley}, {Heger}, $\&$ {Weaver}}{{Woosley et al.}}{2002}]{Woosley2002}
{Woosley} S.E., {Heger} A., $\&$ {Weaver} T.A. 2002, Rev. Mod. Phys., 74, 1015

\bibitem[\protect\citeauthoryear{{Yaron} $\&$ {Gal-Yam}}{{Yaron $\&$ Gal-Yam}}{2012}]{Yaron2012}
{Yaron} O., $\&$ {Gal-Yam} A. 2012, PASP, 124, 668Y

\bibitem[\protect\citeauthoryear{{Zampieri}, {Shapiro}, {Colpi}}{{Zampieri et al.}}{1998}]{Zampieri1998}
{Zampieri} L., {Shapiro} S.L., {Colpi} M., et al. 1998, ApJ, 502L, 149Z

\bibitem[\protect\citeauthoryear{{Zampieri}, {Pastorello}, {Turatto} et al.}{{Zampieri et al.}}{2003}]{Zampieri2003}
{Zampieri} L., {Pastorello} A., {Turatto} M., et al. 2003, \mnras, 338, 711

\bibitem[\protect\citeauthoryear{{Zampieri}}{{Zampieri}}{2005}]{Zampieri2005}
{Zampieri} L., 2005, ASPC, 342, 358Z

\bibitem[\protect\citeauthoryear{{Zampieri}}{{Zampieri}}{2007}]{Zampieri2007}
{Zampieri} L., 2007, AIPC, 924, 358Z

\end{thebibliography}
\end{document}